\documentclass[%
 reprint,
 amsmath,amssymb,
 aps,
 prd,
]{revtex4-2}
\usepackage{import}
\usepackage{dcolumn}
\usepackage{amsmath}  
\usepackage{amsfonts} 
\usepackage{textcomp}
\usepackage[utf8]{inputenc} 
\usepackage[T1]{fontenc}    
\usepackage{hyperref}       
\usepackage{url}            
\usepackage{booktabs}       
\usepackage{nicefrac}       
\usepackage{microtype}      
\usepackage{lipsum}
\usepackage[pdftex]{graphicx}
\usepackage{caption}
\usepackage{amssymb, mathtools, amsthm}
\usepackage[dvipsnames]{xcolor}
\usepackage{mathrsfs}
\usepackage{tensor}
\usepackage{todonotes}
\usepackage{float} 
\usepackage{ragged2e}
\usepackage{scalerel}
\usepackage[normalem]{ulem}
\usepackage[english]{babel}
\usepackage{graphicx}
\usepackage{dcolumn}
\usepackage{bm}
\usepackage{blindtext}
\usepackage{verbatim}
\usepackage{relsize}
\usepackage{musicography}
\usepackage{blindtext}
\usepackage{cancel}
\usepackage{physics}
\usepackage{epstopdf}
\usepackage{mathtools}
\usepackage{blindtext}
\usepackage{color}
\usepackage{epsfig}
\usepackage{pst-grad} 
\usepackage{pst-plot} 
\usepackage{hyperref}
\usepackage{verbatim}
\usepackage{slashed}
\usepackage{subcaption}
\usepackage{dsfont}
\usepackage[english]{babel}
\usepackage{cleveref}

\newcommand{\p}{\partial}

\newcommand{\ZZ}{\mathbb{Z}}
\newcommand{\CC}{\mathbb{C}}
\newcommand{\RR}{\mathbb{{R}}}
\newcommand{\NN}{\mathbb{N}}
\newcommand{\FF}{\mathcal{F}}
\newcommand{\WW}{\mathcal{W}}
\newcommand{\GG}{\mathcal{G}}

\newcommand{\PP}{\mathrm{{P}}}

\newcommand{\Kmax}{K_{\mrm{max}}}
\newcommand{\Kmin}{K_{\mrm{min}}}

\newcommand{\ii}{\mathrm{i}}
\newcommand{\sinc}{\mathrm{sinc}}
\newcommand{\ee}{\mathrm{e}}
\newcommand{\xx}{\mathsf{x}}

\usepackage{float}
\usepackage{tikz-cd}
\usepackage{amsthm}
\usepackage{geometry}
\usepackage{braket}

\DeclareMathOperator{\supp}{supp}

\hypersetup{  colorlinks,
              linktoc         = page, 
              urlcolor        = {green!50!blue},
              linkcolor       = {Bittersweet!80!black}, 
              urlcolor        = {Bittersweet!80!black}, 
              citecolor       = {cyan!50!black}, 
              anchorcolor     = {yellow}
}

\numberwithin{equation}{section}

\theoremstyle{plain}

\theoremstyle{definition}

\newcommand{\mrm}[1]{\mathrm{#1}}
\newcommand{\mc}[1]{\mathcal{#1}}

\newcommand{\lr}[1]{\!\left(#1\right)}
\newcommand{\lrc}[1]{\!\left\{#1\right\}}
\newcommand{\lrb}[1]{\!\left[#1\right]}

\usepackage[final]{showlabels}

\newcommand{\td}[2]{\ensuremath{\frac{\dd #1}{\dd #2}}}

\newcommand{\pd}[2]{\ensuremath{\frac{\partial #1}{\partial #2}}}
\newcommand{\pdn}[3]{\ensuremath{\frac{\partial^{#3} #1}{\partial #2^{#3}}}}

\newcommand{\beq}{\begin{equation}}
\newcommand{\eeq}{\end{equation}}

\newcommand{\ba}{\begin{align}}
\newcommand{\ea}{\end{align}}

\allowdisplaybreaks[1] 

\begin{document}

\title{Circular acceleration in (anti-)de Sitter spacetime: thermality versus finite size}

\author{Cameron R D Bunney}
\email{cameron.r.d.bunney@gmail.com}
\affiliation{School of Mathematical Sciences and School of Physics and Astronomy, 
University of Nottingham, University Park, Nottingham, NG7 2RD, UK}

\author{Jorma Louko}
\email{jorma.louko@nottingham.ac.uk}
\affiliation{School of Mathematical Sciences, University of Nottingham, University Park, Nottingham, NG7 2RD, UK}

\date{June 2024, revised September 2026.}

\begin{abstract}
Anti-de Sitter spacetime and the static patch of de Sitter spacetime are arenas for investigating thermal and finite-size effects seen by an accelerated quantum observer. We consider an Unruh-DeWitt detector in uniform circular motion coupled to a conformal scalar field in $(2+1)$-dimensional de Sitter and anti-de Sitter spacetimes in the limit of a small cosmological constant~$\Lambda$. In anti-de Sitter spacetime, where $\Lambda$ mimics spatial confinement, we find that the resonance peaks in the detector's response closely match those of a detector in Minkowski space with a cylindrical boundary, but with curvature corrections, more significant when the field has an ambient temperature. In the static patch of de Sitter spacetime, in the Euclidean vacuum, we show that the leading curvature correction to the detector's response is proportional to~$\Lambda$, as in zero temperature anti-de Sitter, whilst the temperature corrections decay exponentially as $\Lambda \to 0$.
\end{abstract}

\maketitle

\section{Introduction}


The Unruh effect~\cite{Fulling,Davies1975,Unruh} is one of the most surprising and yet undetected predictions from quantum field theory, stating that a linearly accelerated particle detector with constant proper acceleration $a$ in Minkowski spacetime reacts to a quantum field in its Minkowski vacuum as though the field were in a thermal state in the Unruh temperature $T_U=a\hbar/(2\pi c k_{\mrm{B}})$. The main hurdle in the experimental verification of this effect is the sheer magnitude of acceleration required to reach a detectable increase in temperature.

Unruh-like phenomena exist for non-linear uniformly accelerated motions~\cite{Letaw,Korsbakken:2004bv,Good:2020hav}. One such motion is circular motion, which has long-standing experimental and theoretical interest~\cite{BellLeinaas,Bell:1986ir,Unruh:1998gq,Leinaas:1998tu,ParryDriftRindler}.  A new avenue has been opened by recent proposals~\cite{BEC1,Marino:2020uqj,Gooding,BunneySounds,Gooding:2025tfp} to utilise the analogue spacetime 
that occurs in nonrelativistic laboratory systems \cite{Liberati,HeliumUniverse}. In this paper, we investigate the circular motion Unruh effect in a genuinely relativistic but curved spacetime setting, in $2+1$ de Sitter (dS) and $2+1$ anti-de Sitter (AdS) spacetimes, where positive curvature provides a notion of an ambient temperature, via the Euclidean vacuum, whereas negative curvature provides a notion of spatial confinement. Both of these notions will feature in $(2+1)$-dimensional analogue spacetime realisations of the effect~\cite{Gooding,BunneySounds,Gooding:2025tfp}.
From the broader context, we mention here acceleration effect enhancement by spatial confinement~\cite{CavityUnruh,CavityUnruh2}, 
complex analytic properties of circular motion in (A)dS spacetimes~\cite{AdSdSCircularMotion}, and 
an analogue spacetime simulation proposal of dS spacetime~\cite{BECdeSitter}.

For AdS spacetime, the role of a boundary is played by the asymptotically AdS infinity, where a boundary condition is needed to make the quantum dynamics unitary~\cite{AdSQFT}. We find that in the limit as the cosmological constant tends to zero, an Unruh-DeWitt (UDW) detector probing a scalar field in the global vacuum in AdS reacts to the boundary with resonance peaks closely matching those of a detector in Minkowski spacetime with a cylindrical boundary~\cite{BunneyBoundary}, with subleading terms, which we interpret as curvature corrections. When the scalar field is prepared in a thermal state, however, we find an additional resonance peak with no corresponding term in the response of a detector in Minkowski spacetime with a cylindrical boundary probing a scalar field prepared in a thermal state. 

For the static patch of dS, to which we refer as Rindler-de Sitter spacetime (RdS), the restriction of the Euclidean vacuum is a thermal state~\cite{ChernikovTagirov,BunchDavies,LapedesdeSitter} in a temperature proportional to the square root of the cosmological constant. As the cosmological constant tends to zero, we find that the detector response function is dominated by the contribution from the RdS static vacuum, with the contribution due to thermality exponentially suppressed. This directly parallels the response of a detector probing a scalar field prepared in a thermal state in the low-temperature limit~\cite{BunneyThermal}. We also find the subleading contribution, interpreting it as a curvature correction.

The structure of the paper is as follows. We begin in Section~\ref{sec:: AdS} by establishing the preliminaries for a UDW detector in uniform circular motion in the universal covering space of $(2+1)$-dimensional anti-de Sitter space (CAdS), coupled for a finite time to a quantised, real, massless, conformally coupled scalar field with Dirichlet boundary conditions at spatial infinity, prepared in the global vacuum state. We perform a small-cosmological-constant expansion of the response function to leading and subleading order and interpret the terms in the long-interaction-time limit. In Section~\ref{sec:: thermal ads}, we generalise this analysis to a thermal state, using a derivative-coupled interaction Hamiltonian, which sidesteps the otherwise infrared-divergent thermal Wightman function in the near-Minkowski limit in $(2+1)$ dimensions. We obtain a mode sum expression for the detector's response function and calculate the small-cosmological-constant asymptotic behaviour.

In Section~\ref{sec:: ds}, we perform the corresponding analysis in RdS spacetime with the field in the Euclidean vacuum. The temperature is now determined by the cosmological constant, and the thermal corrections in the detector's response die off exponentially in the near-Minkowski limit, but there is also a curvature correction that decays only proportionally to the cosmological constant.

Section~\ref{sec:: conclusions} provides a summary and concluding remarks. Technical asymptotic expansions are deferred to two appendices.

We use units in which $c=\hbar=k_{\mrm{B}}=1$. Sans serif letters ($\xx$) denote spacetime points and boldface Italic letters ($\bm{k}$) denote spatial vectors. We use metric signature $(-,+,+)$. In asymptotic formulae, $f(x) = O(g(x))$ denotes that $f(x)/g(x)$ remains bounded in the limit considered, $f(x)=o(g(x))$ denotes that $f(x)/g(x)$ tends to zero in the limit considered, and $f(x)\sim g(x)$ denotes that $f(x)/g(x)$ tends to unity in the limit considered.

\section{Anti-de Sitter spacetime}\label{sec:: AdS}
In this Section, we consider the spacetime of constant negative curvature, anti-de Sitter (AdS) spacetime~\cite{HawkingEllis}. AdS is not globally hyperbolic owing to its timelike boundary at spatial infinity, through which information can propagate~\cite{AdSQFT}. However, by imposing boundary conditions at spatial infinity, one can employ a consistent quantisation scheme for a quantum field~\cite{AdSQFT,Ishibashi_Wald,Ambrus:2018olh, Dappiaggi:2018pju,PitelliComment,VitorAdS_BC, Morley_2021, Dappiaggi:2022dwo}. We adopt the Dirichlet boundary condition and consider the response of the detector in the small cosmological constant limit, with the field in its global vacuum. Thermal states will be considered in Section~\ref{sec:: thermal ads}.

\subsection{Spacetime, field, and detector preliminaries}\label{subsec:: prelims}
Anti-de Sitter spacetime in $2+1$ dimensions can be embedded in $\RR^{2,2}$ with the embedding equation
\begin{equation}
    \eta^{2,2}_{AB}X^AX^B~=~-\alpha^2\,,
\end{equation}where $X^A=(T^0,T^1,X^1,X^2)$ are the coordinates on $\RR^{2,2}$, $\eta^{2,2}_{AB}=\text{diag}(-1,-1,1,1)$, and $\alpha>0$ is the radius of curvature~\cite{HawkingEllis}. The cosmological constant $\Lambda$ is related to $\alpha$ by $\Lambda = -1/\alpha^2$. The Ricci scalar of $\mrm{AdS}_3$ is $\mc{R}=-6/\alpha^2$. We have adopted the notation $\eta^{p,q}$ to emphasise the signature of the embedding space.

In this paper, we work in the universal covering space of anti-de Sitter spacetime CAdS, which contains no closed timelike curves. Points on CAdS are denoted by $\xx$. We use the global coordinates $(t,r,\theta)$ in which the metric on $\mrm{AdS}_3$ is given by
\begin{equation}\label{eqn:: AdS3 metric}
    \dd s^2~=~-\lr{1+\frac{r^2}{\alpha^2}}\dd t^2+\frac{\dd r^2}{1+\frac{r^2}{\alpha^2}}+r^2\dd\theta^2\,,
\end{equation}where $-\infty<t<\infty$, $0\leq r<\infty$, and $0\leq\theta<2\pi$, with the usual coordinate singularity at $r=0$ and $\theta$ periodicially identified with period $2\pi$. The embedding in the embedding space $\RR^{2,2}$ is given by
\begin{subequations}\label{eqn:: global ads coords}
    \begin{align}
    T^0&~=~\sqrt{r^2+\alpha^2}\cos(t/\alpha)\,,\\
    T^1&~=~\sqrt{r^2+\alpha^2}\sin(t/\alpha)\,,\\
    X^1&~=~r\cos\theta\,,\\
    X^2&~=~r\sin\theta\,.
\end{align}
\end{subequations}
The small cosmological constant limit $\Lambda\rightarrow 0^-$ is given by the limit $\alpha\rightarrow\infty$, which recovers Minkowski spacetime $\RR^{2,1}$, as can be seen by the form of the metric~\eqref{eqn:: AdS3 metric}.

We consider a quantised, real, massless, conformally coupled scalar field $\Phi$ and we probe the field with a pointlike detector in uniform circular motion,
\begin{equation}\label{eqn:: trajectory}
    (t,r,\theta)~=~(\gamma\tau,R,\Omega\gamma\tau)\,,
\end{equation}where $\tau$ is proper time, $R>0$ is the radius, $\Omega=\frac{\dd \theta}{\dd t}> 0$ is the angular velocity, and
\begin{equation}\label{eqn:: gamma}
    \gamma~=~\frac{1}{\sqrt{1-R^2\Omega^2+\frac{R^2}{\alpha^2}}}\,,
\end{equation}to which we refer as the Lorentz factor for circular motion in AdS. We assume that the worldline is timelike, $R\Omega<\sqrt{1+R^2/\alpha^2}$.

In the large-boundary limit, we have $\lim_{\alpha\rightarrow\infty}\gamma=\frac{1}{\sqrt{1-R^2\Omega^2}}=\Gamma$, where $\Gamma$ is the Lorentz factor associated with circular motion in Minkowski spacetime $\RR^{2,1}$.

The detector's Hilbert space is given by $\mc{H}_\mrm{D}\simeq\CC^2$ and is spanned by the orthonormal basis $\{\ket{0},\ket{1}\}$. The Hamiltonian of the detector $H_\mrm{D}$, generating the dynamics with respect to the proper time $\tau$, acts on $\mc{H}_\mrm{D}$ as $H_\mrm{D}\ket{0}=0$ and $H_\mrm{D}\ket{1}=E\ket{1}$, where $E\in\RR\setminus\{0\}$. The detector, therefore, is a two-level system with an energy gap $E$: for $E>0$, $\ket{0}$ is the ground state and $\ket{1}$ is the excited state; for $E<0$, the roles are reversed.

In the interaction picture, we take the interaction Hamiltonian to be
\begin{equation}\label{eqn:: interaction hamiltonian}
H_\mrm{I}~=~\lambda\chi(\tau)\Phi(\xx(\tau))\otimes\mu(\tau)\,,
\end{equation}where $\mu$ is the detector's monopole moment operator, $\chi$ is a real-valued switching function specifying the temporal profile of the interaction, and $\lambda$ is a real-valued coupling constant, assumed to be small.

Working to first order in perturbation theory in $\lambda$, the probability for the detector to transition from $\ket{0}$ to $\ket{1}$, regardless of the final state of the field, is given by~\cite{Unruh, birrell}
\begin{equation}
    \mc{P}(E)~=~\lambda^2|\braket{1|\mu(0)|0}|^2\mc{F}(E)\,,
\end{equation} where $\mc{F}$ is the \textit{response function}, given by
\begin{equation}\label{eqn:: response}
    \mc{F}(E)~=~\int\dd \tau'\dd \tau''\chi(\tau')\chi(\tau'')\ee^{-\ii E(\tau'-\tau'')}\mc{W}(\tau',\tau'')\,,
\end{equation} and $\mc{W}(\tau',\tau'')$ is the pullback of the Wightman function $\mc{W}(\xx',\xx'')=\braket{\Phi(\xx')\Phi(\xx'')}$ to the circular trajectory.

We prepare the field $\Phi$ in the global vacuum state, defined with respect to the global timelike Killing vector $\p_t$ in coordinates~\eqref{eqn:: AdS3 metric}. Then, the Wightman function for a scalar field in $\mrm{AdS}_3$ is given by~\cite{OrtizBH,LoukoBTZ}
\begin{subequations}
    \begin{align}\label{eqn:: Wightman AdS}
    \mc{W}(\xx',\xx'')&~=~\frac{1}{4\pi}\lr{\frac{1}{\sqrt{\sigma(
    \xx',\xx''
    )}}+\frac{\zeta}{\sqrt{\sigma(\xx',\xx'')+4\alpha^2}}}\,,\\
    \sigma(\xx',\xx'')&~=~\eta^{2,2}_{AB}\lr{{X'}^A-{X''}^A}\lr{{X'}^B-{X''}^B}\,,\label{eqn:: sigma}
\end{align}
\end{subequations}
where $\zeta\in\{-1,0,1\}$ corresponding to Dirichlet, transparent, or Neumann boundary conditions at spatial infinity and $\sigma$ is the geodesic squared distance in the embedding space. We will only consider Dirichlet boundary conditions, $\zeta=-1$. We adopt the notation $\mc{W}(\tau',\tau'')=\mc{W}(\xx(\tau'),\xx(\tau''))$ and $\sigma(\tau',\tau'')=\sigma(\xx(\tau'),\xx(\tau''))$ for the pullback of the Wightman function~\eqref{eqn:: Wightman AdS} and geodesic squared distance~\eqref{eqn:: sigma} to a trajectory $\xx(\tau)$ parametrised by its proper time. Along a stationary trajectory~\cite{Letaw,LetawPfautsch,BunneyStationary}, such as uniform circular motion, the pullback of the Wightman function is time-translation invariant $\mc{W}(\tau',\tau'')=\mc{W}(\tau'-\tau'',0)$. The Wightman function should be understood in the distributional sense as $\lim_{\varepsilon\to0^+}\mc{W}(\tau'-\tau''-\ii\varepsilon,0)$. Then, in~\eqref{eqn:: Wightman AdS}, the branch of the square root is taken such that the limit $\varepsilon\to0^+$ of the square root is positive when $\sigma(\xx',\xx'')>0$~\cite{LoukoBTZ}.

The geodesic squared distance in global coordinates along the circular trajectory is given by
\begin{multline}
    \label{eqn:: ads sigma}
    \sigma(s,0)~=~\\-4\alpha^2\sin^2\lr{\frac{\gamma s}{2\alpha}}-2R^2\lr{\cos(\Omega\gamma s)-\cos\lr{\frac{\gamma s}{\alpha}}}\,.
\end{multline}
The Wightman function associated with~\eqref{eqn:: ads sigma} has a pole at $s=0$ and two infinite families of branch points. In the large-boundary limit, these branch points lie around $s=\frac{2\alpha}{\gamma}\pi n$ and $s=\frac{2\alpha}{\gamma}\frac{(2k+1)}{2}\pi$ with $n,~k\in\ZZ$. As such, the singularity structure of the Wightman function in $\mrm{CAdS}_3$ is fundamentally different from the Wightman function of the Minkowski vacuum in $\RR^{2,1}$, which has a single pole at $s=0$.

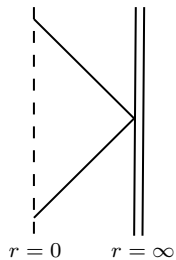
\begin{figure}[t]
\centering{\tikzset{every picture/.style={line width=0.75pt}} 

\begin{tikzpicture}[x=0.75pt,y=0.75pt,yscale=-1,xscale=1]

\draw    (150.83,175) -- (150,290) ;
\draw  [dash pattern={on 4.5pt off 4.5pt}]  (100,175) -- (100,290) ;
\draw    (155,175) -- (154.17,290) ;
\draw    (100,281) -- (150,231) ;
\draw    (150,231) -- (100,181) ;

\draw (86,292.4) node [anchor=north west][inner sep=0.75pt]  [font=\footnotesize]  {$r=0$};
\draw (137,294.7) node [anchor=north west][inner sep=0.75pt]  [font=\footnotesize]  {$r=\infty $};

\end{tikzpicture}}
\caption{\label{fig:: conformal_diagram}\justifying Conformal diagram of CAdS.}
\end{figure}

\subsection{Finite-time interaction}\label{subsec:: finite interaciton time}
To sidestep the technical challenge of handling the branch points directly, we use a switching function $\chi$ of compact support. For $\alpha$ sufficiently large, the only singularity in the domain of the integral is the pole at $s=0$. Keeping the support of the switching function fixed, we perform a large-$\alpha$ expansion under the integral by a dominated-convergence argument. After this, we take the long-interaction-time limit.

Using the time-translation of the Wightman function along the circular trajectory~\eqref{eqn:: ads sigma}~\eqref{eqn:: Wightman AdS}, we rewrite the response function~\eqref{eqn:: response} as
\begin{equation}\label{eqn:: response X}
    \FF(E)~=~\int\dd s\,X(s)\ee^{-\ii Es}\mc{W}(s,0)\,,
\end{equation} where $X$ is the self-convolution of $\chi$,
\begin{equation}\label{eqn:: self convolution}
X(s)~=~\int\dd\tau'\,\chi(\tau'+\tfrac{1}{2}s)\chi(\tau'-\tfrac{1}{2}s)\,.
\end{equation}We assume $\chi$ to be smooth and of compact support. $X$ is then also smooth and of compact support. In the long-interaction-time limit, one usually lets $\chi\to1$ and divides by the total interaction time because the response function~\eqref{eqn:: response} diverges linearly with the interaction time. In terms of $X$~\eqref{eqn:: self convolution}, this process amounts to the limit $X\to1$.

In $\FF(E)$~\eqref{eqn:: response X}, we first handle the singular behaviour at $s=0$ coming from the term $1/(4\pi\sqrt{\sigma(s,0)}\,)$ in~\eqref{eqn:: Wightman AdS}, understood again in the distributional sense $\lim_{\varepsilon\rightarrow0^+}1/(4\pi\sqrt{\sigma(s-\ii\varepsilon,0)}\,)$, where the square root in the denominator is positive imaginary for $s>0$ and negative imaginary for $s<0$.

Around $s=0$, we have the expansion
\begin{equation}\label{eqn:: Wightman small s}
    \mc{W}(s-\ii\varepsilon,0)\sim\frac{1}{4\pi \ii(s-\ii\varepsilon)}\,.
\end{equation}We can isolate this singular contribution to the Wightman function by adding and subtracting it~\cite{Biermann,BunneyRick}
\begin{multline}\label{eqn:: F sing split}
    \FF(E)~=~\int\dd s\,X(s)\ee^{-\ii Es}\\
    \times\lr{\mc{W}(s-\ii\varepsilon,0)-\frac{1}{4\pi \ii(s-\ii\varepsilon)}+\frac{1}{4\pi \ii(s-\ii\varepsilon)}}\,.
\end{multline}In the first two terms in~\eqref{eqn:: F sing split}, the singularities at $s=0$ cancel, and the $\varepsilon\to0^+$ limit can be taken under the integral. We may hence split the integral as
\begin{subequations}\label{eqn:: response f0}
\begin{align}
    \FF(E)~=&~\FF_0(E)+\FF_{\text{corr}}(E)\,,\\
    \label{eqn:: sing part}
    \FF_0(E)~=&~\int\dd s\,\frac{X(s)\ee^{-\ii Es}}{4\pi\ii(s-\ii\varepsilon)}\,,\\
    \FF_{\text{corr}}(E)~=&~\int\dd s\,X(s)\ee^{-\ii Es}\lr{\mc{W}(s,0)-\frac{1}{4\pi\ii s}}\,.\label{eqn:Fcorr}
\end{align}\end{subequations} 
To~\eqref{eqn:: sing part}, we use the Sokhotski-Plemelj theorem~\cite{TitchmarshSokhotski},
\begin{equation}
    \lim_{\varepsilon\rightarrow0^+}\int_a^b\dd x\,\frac{f(x)}{x-\ii\varepsilon}=\ii\pi f(0)+\mc{P}\int_a^b\dd x\,\frac{f(x)}{x}\,,
\end{equation} where $\mc{P}$ indicates the Cauchy principal value. Applying this to~\eqref{eqn:: sing part}, we have
\begin{subequations}\label{eqn:: f0}
    \begin{equation}
   \FF_0(E)~=~\frac{1}{4}X(0)-\frac{1}{2\pi}\int_0^\infty\dd s\,\frac{X(s)\sin(Es)}{s}\,,
\end{equation}
\end{subequations}
where we have used that $X(s)$ is even.

Owing to the compact support of $X$, we can express the integrand of $\FF_{\text{corr}}$~\eqref{eqn:Fcorr} as an asymptotic power series in $\alpha$, where for sufficiently large $\alpha$ the singularities of the Wightman function are outside of the support of $X$ and the error term is uniform in $s$. Using~\eqref{eqn:: Wightman AdS} with $\zeta=-1$, $\FF_{\text{corr}}$~\eqref{eqn:Fcorr} can be written as
\begin{multline}\label{eqn:: f corr split}
        \FF_{\text{corr}}(E)~=~\frac{1}{4\pi\ii}\int\dd s\,X(s)\ee^{-\ii Es}\lr{\frac{1}{\sqrt{-\sigma(s,0)}}-\frac{1}{s}}\\
        -\frac{1}{8\pi\alpha}\int\dd s\,X(s)\ee^{-\ii Es}\frac{1}{\sqrt{1+\frac{\sigma(s,0)}{4\alpha^2}}}\,.
\end{multline}
We recall that  $\sqrt{-\sigma(s,0)}$ is positive for $s>0$ and negative for $s<0$, the integrand in the first term in~\eqref{eqn:: f corr split} is nonsingular at $s=0$, $\sigma(s,0)$~\eqref{eqn:: ads sigma} for fixed $s\neq0$ has a large-$\alpha$ expansion in integer powers of $\alpha^{-2}$, and $X$ has compact support. It follows that the large-$\alpha$ expansion of~\eqref{eqn:: f corr split} proceeds in integer powers of $\alpha^{-1}$. The even powers will come from the first term and the odd powers will come from the second term. Note that the first term is independent of the boundary condition at infinity in~\eqref{eqn:: Wightman AdS} and so contains only curvature corrections, whereas the second term is responsible for both boundary effects and curvature corrections.

\subsection{Large-$\alpha$ limit}\label{subsec:: large boundary zero}
In this Section, we calculate the first four terms in the large-$\alpha$ expansion of the response function~\eqref{eqn:: response f0}. We interpret each term in the long-interaction-time limit and compare these terms with the large-boundary limit in $2+1$ Minkowski spacetime with a cylindrical boundary~\cite{BunneyBoundary}.

We begin with the terms of order unity. The first is given by $\FF_0$~\eqref{eqn:: f0} and the second is found by a large-$\alpha$ expansion under the first integral in~\eqref{eqn:: f corr split},
\begin{subequations}
    \begin{align}
    \FF_{\text{corr}}^{(0)}(E)~=&~\lim_{\alpha\rightarrow\infty}\FF_{\text{corr}}(E)\\\nonumber~=&~\frac{1}{2\pi}\int_0^\infty\dd s\,X(s)\frac{\sin(Es)}{s}\\&\times\lr{1-\frac{1}{\sqrt{\Gamma^2-4R^2\sin^2\lr{\tfrac{\Gamma\Omega}{2}s}/s^2}}}\,,\label{eqn:: ads large a}
\end{align}
\end{subequations}where $\Gamma=(1-R^2\Omega^2)^{-1/2}$ is the Lorentz factor associated with circular motion in Minkowski spacetime. We are able to take the limit under the integral by the dominated convergence theorem owing to the compact support of $X$. Since the integrand of~\eqref{eqn:: ads large a} is even, we have combined the $s<0$ and $s>0$ contributions to write the integral over $s>0$.

We find the order-$\alpha^{-1}$ contribution by a large-$\alpha$ expansion under the second integral in~\eqref{eqn:: f corr split}, leading to
\begin{equation}\label{eqn:: f alpha 1}
    \FF_{\text{corr}}^{(1)}~=~-\frac{1}{8\pi\alpha}\int\dd s\,X(s)\ee^{-\ii Es}\,.
\end{equation}

The order-$\alpha^{-2}$ contribution comes from the large-$\alpha$ under the first integral in~\eqref{eqn:: f corr split}
\begin{multline}\label{eqn:: f2 ads}
    \FF_{\text{corr}}^{(2)}(E)~=~\\-\frac{1}{8\pi\ii\alpha^2}\int\dd s\,\frac{X(s)\ee^{-\ii Es}}{\lr{\Gamma^2-4R^2\sin^2\lr{\tfrac{\Gamma\Omega}{2}s}/s^2}^{3/2}}\\
    \times\lr{-\frac{1}{12}\Gamma^4s+R^4\Omega\Gamma^3\frac{\sin(\Omega\Gamma s)-\Omega\Gamma s}{s^2}}\,.
\end{multline}
The integrand is nonsingular at $s=0$.

The order-$\alpha^{-3}$ contribution arises from the large-$\alpha$ expansion under the second integral in~\eqref{eqn:: f corr split},
\begin{multline}\label{eqn:: f3 alpha}
    \FF_{\text{corr}}^{(3)}(E)~=~-\frac{1}{64\pi\alpha^3}\int\dd s\,X(s)\ee^{-\ii Es}\\\times\lr{\Gamma^2s^2-4R^2\sin^2\lr{\tfrac{\Gamma\Omega}{2}s}}\,.
\end{multline}

\subsubsection{Long-interaction-time limit}
We consider now the long-interaction-time limit $X(s)\to1$ of the terms in the asymptotic expansion. The leading-order contributions are the limiting forms of~\eqref{eqn:: f0} and~\eqref{eqn:: ads large a},
\begin{subequations}\label{eqn: F0 long X}
    \begin{align}
    \FF_0(E)&~=~\frac{1}{2}\Theta(-E)\,,\\
    \FF^{(0)}_{\mrm{corr}}(E)&~=~\frac{1}{2\pi}\int_0^\infty\dd s\,\frac{\sin(Es)}{s}\nonumber\\
    &\times\lr{1-\frac{1}{\sqrt{\Gamma^2-4R^2
    \sin^2\lr{\tfrac{\Gamma\Omega}{2}s}/s^2}}}\,,
\end{align}
\end{subequations}which is the response function for a detector in uniform circular motion on a trajectory with radius $R$ and angular velocity $\Omega$ in $(2+1)$ Minkowski spacetime~\cite{Biermann}.

The order-$\alpha^{-1}$ term~\eqref{eqn:: f alpha 1} in the long-interaction-time limit is
\begin{equation}
    \FF_{\text{corr}}^{(1)}(E)~=~-\frac{1}{4\alpha}\delta(E)\,,
\end{equation}which vanishes as $E\neq0$ by assumption.

The order-$\alpha^{-2}$ term~\eqref{eqn:: f2 ads} is
\begin{multline}\label{eqn: F2 long X}
    \FF_{\text{corr}}^{(2)}(E)~=~\\-\frac{1}{4\pi\alpha^2}\int_0^\infty\dd s\,\frac{\sin(Es)}{{\lr{\Gamma^2-4R^2\sin^2\lr{\tfrac{\Gamma\Omega}{2}s}/s^2}}^{3/2}}\\
    \times\lr{-\frac{1}{12}\Gamma^4s+R^4\Omega\Gamma^3\frac{\sin(\Omega\Gamma s)-\Omega\Gamma s}{s^2}}\,.
\end{multline}We shall return to the interpretation of this term at the end of this Section.

The order-$\alpha^{-3}$ term~\eqref{eqn:: f3 alpha} can be expressed as 
\begin{align}
    \FF_{\text{corr}}^{(3)}(E)~=~&\frac{1}{64\pi\alpha^3}\lr{2R^2+\Gamma^2\frac{\p^2}{\p E^2}}\int\dd s\,X(s)\ee^{-\ii Es}\nonumber\\
    -\frac{R^2}{64\pi\alpha^3}&\int\dd s\,X(s)\lr{\ee^{-\ii(E-\Gamma\Omega)s}+\ee^{-\ii(E+\Gamma\Omega)s}}\,.\label{eqn:: f3}
\end{align}In the long-interaction-time limit, the first term in~\eqref{eqn:: f3} has the distributional limit
\begin{equation}
    \frac{1}{32\alpha^3}\lr{2R^2+\Gamma^2\pdn{}{E}{2}}\delta(E)\,,
\end{equation}which vanishes since $E\neq0$ by assumption. The second term in~\eqref{eqn:: f3} has the limit
\begin{equation}\label{eqn:: f3 delta}
    \FF_{\text{corr}}^{(3)}(E)~=~-\frac{R^2}{32\alpha^3}\lr{\delta(E-\Gamma\Omega)+\delta(E+\Gamma\Omega)}\,.
\end{equation} The deltas in~\eqref{eqn:: f3 alpha} are resonance peaks at the detector's angular frequency.

Recall that we introduced the anti-de Sitter length scale $\alpha$ as a model of a finite-size cylindrical boundary in Minkowski spacetime, which is investigated in~\cite{BunneyBoundary}, and we may now compare our expansion~\eqref{eqn: F0 long X}-~\eqref{eqn:: f3 delta} to that found in~\cite{BunneyBoundary}. Our order-$\alpha^0$ contribution~\eqref{eqn: F0 long X} is equal to the Minkowski spacetime response function. Our next contribution~\eqref{eqn: F2 long X} is order $\alpha^{-2}$ and the response in Minkowski spacetime with cylindrical boundary has no corresponding term. This term must hence be interpreted as a curvature correction. Indeed, the Ricci scalar of $\mrm{CAdS}_3$ is proportional to $\alpha^{-2}$. Our next contribution~\eqref{eqn:: f3 delta} is order~$\alpha^{-3}$, coming from the boundary contribution to the Wightman function~\eqref{eqn:: Wightman AdS}, and we find that this term does agree, in its falloff and its resonance structure, with the corresponding term in Minkowski spacetime with a cylindrical boundary.

\subsection{Static observers}
In this Section, we consider the term-wise behaviour of the response function for a static observer, obtained by setting $\Omega=0$. The proper acceleration of a static observer is given by
\begin{equation}
    A~=~\frac{1}{\alpha}\frac{\lr{\frac{R}{\alpha}}}{\sqrt{1+\lr{\frac{R}{\alpha}}^2}}\,,
\end{equation}
which is bounded above by $\alpha^{-1}$ and reduces to $0$ for the geodesic observer at $R=0$. The limit $\lim_{\Omega\to0}\FF(E)$ is well defined both at the level of the full response function and also at the level of the asymptotic expansion in $\alpha$~\eqref{eqn:: ads large a}-\eqref{eqn:: f3 alpha}; hence, we simply evaluate the terms in this expansion at $\Omega=0$, obtaining
\begin{subequations}\label{eqn:: static expansion}
    \begin{align}
    \left.\FF_0(E)\right|_{\Omega=0}&~=~\FF_0(E)\,,\\
    \left.\FF^{(0)}_{\text{corr}}(E)\right|_{\Omega=0}&~=~0\,,\\
    \left.\FF^{(1)}_{\text{corr}}(E)\right|_{\Omega=0}&~=~-\frac{1}{8\pi\alpha}\int\dd s\, X(s)\cos(Es)\,,\\
    \left.\FF^{(2)}_{\text{corr}}(E)\right|_{\Omega=0}&~=~-\frac{1}{96\pi\alpha^2}\int\dd s\,sX(s)\sin(Es)\,,\\
    \left.\FF^{(3)}_{\text{corr}}(E)\right|_{\Omega=0}&~=~-\frac{1}{64\pi\alpha^3}\int\dd s\,s^2X(s)\cos(Es)\,,
\end{align}
\end{subequations}where we have used the evenness of $X(s)$. None of the above terms depend on $R$; however, it can be shown that the order-$\alpha^{-4}$ contribution $\FF^{(4)}_{\text{corr}}(E)$, not displayed here, does.

In the long-interaction-time limit, $X(s)\to1$, Eqs.~\eqref{eqn:: static expansion} reduce to
\begin{subequations}
\begin{align}
    \left.\FF_0(E)\right|_{\Omega=0}&~=~\FF_0(E)\,,\\
    \left.\FF^{(0)}_{\text{corr}}(E)\right|_{\Omega=0}&~=~0\,,\\
    \left.\FF^{(1)}_{\text{corr}}(E)\right|_{\Omega=0}&~=~-\frac{1}{4\alpha}\delta(E)\,,\label{eqn:: f1 corr delta}\\
    \left.\FF^{(2)}_{\text{corr}}(E)\right|_{\Omega=0}&~=~\frac{1}{24\alpha^2}\pd{}{E}\delta(E)\,,\\
    \left.\FF^{(3)}_{\text{corr}}(E)\right|_{\Omega=0}&~=~\frac{1}{16\alpha^3}\pdn{}{E}{2}\delta(E)\,,\label{eqn:: f3 corr delta}
\end{align}\end{subequations}obtained by writing the powers of $s$ under the integral in~\eqref{eqn:: static expansion} as derivatives with respect to $E$ outside the integral. Since we have assumed $E\neq0$, the correction terms~\eqref{eqn:: f1 corr delta}-\eqref{eqn:: f3 corr delta} vanish. Proceeding to higher orders in $\alpha$, all correction terms in this limit are derivatives of $\delta(E)$ and vanish. This is consistent with the expectation from the Global Embedding Minkowski Spacetime (GEMS) paradigm~\cite{Deser_1997, Jacobson_Deser_Reply} that a detector on the static worldline, with acceleration less than $\alpha^{-1}$, should not see thermality, and with the static detector analysis in~\cite{Jennings_AdS}.

\section{Thermal states in CAdS}\label{sec:: thermal ads}
In this Section, we generalise the anti-de Sitter analysis of Section \ref{sec:: AdS} from the global vacuum to a thermal state in inverse temperature~$\beta$. We first consider thermal states in a general static spacetime and then specialise to CAdS. We calculate the leading corrections to the response function in the large-$\alpha$ limit.

\subsection{Thermal states in static spacetimes}\label{sec:thermal states}
We consider a conformally coupled, massless, real, scalar field prepared in a thermal state in an $(n+1)$-dimensional static spacetime $(\mc{M},g_{\mu\nu})$. Such a spacetime admits a coordinate system in which the metric is given by
\begin{equation}\label{eqn:: static metric}
    \dd s^2~=~-g_{00}(\bm{x})\dd t^2 + h_{ij}(\bm{x})\dd x^i \dd x^j\,.
\end{equation} The conformally coupled, massless, Klein-Gordon equation is given by
\begin{subequations}
    \begin{align}
    0&~=~\lr{\Box-\xi R}\Phi(\xx)\,\\
   & ~=~\Biggl( g^{00}\nonumber(\bm{x})\p_0\p_0\\&\left.+\frac{1}{\sqrt{-g(\bm{x})}}\p_i\sqrt{-g(\bm{x})}h^{ij}(\bm{x})\p_j-\xi(n) \mc{R}\right)\Phi(\xx)\,,\label{eqn:: KG static}
\end{align}
\end{subequations}where $g(\bm{x})=\det(g_{\mu\nu})$, $\mc{R}$ is the Ricci scalar curvature, and $\xi(n)=(n-1)/(4n)$ is the conformal coupling. If we consider the mode decomposition
\begin{equation}
    \Phi(\xx)~=~\sum_k\phi_k(\xx)a_k+\phi_k^*(\xx)a^\dagger_k\,,
\end{equation}then the field modes $\phi_k$ are also solutions to~\eqref{eqn:: KG static}. We will specify the momenta $k$ in more detail later. Due to the separation time and space in~\eqref{eqn:: KG static}, the equation admits a separable solution. The positive-frequency field modes with respect to the Killing vector $\p_t$ are of the form
\begin{equation}\label{eqn:: static modes}
    \phi_k(\xx)~=~\frac{1}{\sqrt{2\omega_k}}f_k(\bm{x})\ee^{-\ii \omega_k t}\,.
\end{equation}In static spacetimes in coordinates adapted to the staticity~\eqref{eqn:: static metric}, the Klein-Gordon norm is given by
\begin{equation}\label{eqn:: KG norm}
    (\phi_1,\phi_2)\!=\ii\int_\Sigma\dd^n \bm{x}\,\sqrt{-g(\bm{x})}g^{00}(\bm{x})\!\lr{\phi_1\p_0\phi_2^*-\phi_2^*\p_0\phi_1}\,,
\end{equation}where $\Sigma$ is any Cauchy surface. We assume that $\{\phi_k\}$ forms a complete set, which we normalise with respect to the Klein-Gordon norm~\eqref{eqn:: KG norm},
\begin{align}\label{eqn:: KG normalisation}
    (\phi_k(\xx),\phi_{k'}(\xx))&~=~\delta_{kk'}~=~-(\phi^*_k(\xx),\phi^*_{k'}(\xx))\,,\\
     (\phi_k(\xx),\phi^*_{k'}(\xx))&~=~0\,.
\end{align}As a consequence of this normalisation~\eqref{eqn:: KG normalisation}, the completeness relation of the spatial modes $f_k$ is
\begin{equation}
    -\int\dd^{n}\bm{x}\,\sqrt{-g(\bm{x})}g^{00}(\bm{x})f_k(\bm{x})f^*_{k'}(\bm{x})~=~\delta_{kk'}\,.
\end{equation}

By restricting to a static spacetime, the field Hamiltonian $H$ is time independent and for field modes~\eqref{eqn:: static metric} $H$ is given by~\cite{DowkerTemp,HamiltonianDiagonalisation,StaticSpaceTimeHamiltonian,ESUHamiltonian}
\begin{equation}\label{eqn:: Hamiltonian}
    H~=~\frac{1}{2}\sum_k \omega_k\lr{a_ka_k^\dagger+a_k^\dagger a_k}\,.
\end{equation}

Due to the time independence of the Hamiltonian~\eqref{eqn:: Hamiltonian}, conventional concepts about thermal averaging are still valid in this spacetime~\cite{DowkerTemp}. The thermal expectation value of an operator $A$ at inverse temperature $\beta=1/T$ is given by the statistical average over a canonical ensemble,
\begin{equation}\label{eqn:: thermal expectation}
    \braket{A}_\beta~=~\frac{\Tr\lr{\ee^{-\beta H}A}}{\Tr\lr{\ee^{-\beta H}}}\,,
\end{equation} 
and the local temperature observed by a static observer on a (static) curved spacetime is given by Tolman's law~\cite{TolmanPrelim,TolmanEhrenfest,TolmanBook},
\begin{equation}
    T_{\mrm{obs}}(\bm{x})~=~\frac{T}{\sqrt{-g_{00}(\bm{x})}}\,.
\end{equation}In anti-de Sitter spacetime in the coordinates~\eqref{eqn:: global ads coords}, we have $-g_{00}(\bm{x})=1$ at the origin, giving a distinguished reference point. 
In the Rindler-de Sitter spacetime of Section~\ref{sec:: ds}, we will use the coordinates~\eqref{eqn:: ds metric}, in which the origin is similarly distinguished. In writing $\exp(-\beta H)$ in~\eqref{eqn:: thermal expectation}, we consider $\beta$ as measured by a static observer at the origin.

The operator in whose thermal expectation value we are interested is $\Phi(\xx)\Phi(\xx')$. This operator-valued distribution is well defined and requires no renormalisation. It follows from~\eqref{eqn:: thermal expectation} that the thermal expectation value is given by
\begin{multline}\label{eqn:: therm exp}
\braket{\Phi(\xx)\Phi(\xx')}_\beta~=~\sum_k(1+n(\beta\omega_k))\phi_k(\xx)\phi^*_k(\xx')\\+n(\beta\omega_k)\phi^*_k(\xx)\phi_k(\xx')\,,
\end{multline}where $n(x)=(\ee^x-1)^{-1}$ is the Bose thermal factor. The distributional character of $\braket{\Phi(\xx)\Phi(\xx')}_\beta$ can be encoded by the insertion of suitable $\ii\varepsilon$ convergence factors in~\eqref{eqn:: therm exp}.

\subsection{Field and detector preliminaries}\label{sec:: thermal prelims}
Specialising now to CAdS, we consider a quantised, conformally coupled, massless scalar field $\Phi$,
\begin{equation}
    \Phi(t,r,\theta)~=~\sum_{k}\phi_{k}(t,r,\theta)a_{k}+\text{H.c.}\,,
\end{equation}
where the annihilation $a_{k}$ and creation $a^\dagger_{k}$ operators obey the commutation relation $[a_{k},a^\dagger_{k'}]=\delta_{k,k'}$. The Hilbert space is a Fock space with a Fock vacuum $\ket{0}$ satisfying $a_{k}\ket{0}$ for all~$k$.

We prepare the field in a thermal state in inverse temperature~$\beta$. In place of~\eqref{eqn:: interaction hamiltonian}, we use a derivative-coupled interaction Hamiltonian~\cite{BunneyThermal,BunneyBoundary},
\begin{equation}\label{eqn:: deriv interact}
    H_{\mrm{I}}~=~\lambda\chi(\tau)\frac{\dd}{\dd \tau}\Phi(\xx(\tau))\otimes\mu(\tau)\,.
\end{equation}The derivative in~\eqref{eqn:: deriv interact} cures the infrared divergence that occurs in the finite temperature Wightman function in $2+1$ dimensions in the $\alpha\to\infty$ limit~\cite{BunneyThermal,BunneyBoundary}. In higher spacetime dimensions the thermal Wightman function is no longer infrared divergent in this limit, and using the Hamiltonian~\eqref{eqn:: deriv interact} is equivalent to using the Hamiltonian~\eqref{eqn:: interaction hamiltonian} multiplied by $E$~\cite{RickDerivative}.

Working again to first order in perturbation theory in $\lambda$, the thermal response function is given by
\begin{equation}\label{eqn:: therm response}
    \FF(E,\beta)=\!\!\int\dd\tau'\dd\tau''\,\chi(\tau')\chi(\tau'')\ee^{-\ii E(
    \tau'-\tau'')}\mc{W}_\beta(\tau',\tau'')\,,
\end{equation}where $\mc{W}_\beta(\tau',\tau'')=\braket{\tfrac{\dd}{\dd\tau'}\Phi(\xx(\tau'))\tfrac{\dd}{\dd\tau''}\Phi(\xx(\tau''))}_\beta$ is the thermal Wightman function.

We can calculate the thermal Wightman function in the same way as the vacuum Wightman function construction in~\cite{OrtizBH} by utilising that CAdS is conformal to half of Einstein Static Universe (ESU)~\cite{AdSQFT}. Then, the field modes in CAdS, normalised with respect to the Klein-Gordon norm~\eqref{eqn:: KG norm}, are given by
\begin{multline}
    \phi_{lm}(t,r,\theta)~=~\frac{1}{\lr{1+\frac{r^2}{\alpha^2}}^{1/4}}\sqrt{\frac{1}{4\pi\alpha}\frac{(l-m)!}{(l+m)!}}\\ \times \PP_l^m\lr{\frac{1}{\sqrt{{1+\frac{r^2}{\alpha^2}}}}}\ee^{\ii m\theta-\ii\omega_l t}\,,
\end{multline}where $l\in\NN_0=\{0,1,2,\dots\}$, $|m|\leq l$, $\omega_l=\tfrac{1}{\alpha}(l+\tfrac{1}{2})$, and $\PP^m_l$ is the associated Legendre function of degree $l$ and order $m$, with argument taking values between $-1$ and $+1$, also known as Ferrers' functions or associated Legendre functions \textit{on the cut}~\cite{NIST}. Appropriate boundary conditions are required to ensure the well-posedness of the Cauchy problem~\cite{AdSQFT, Ishibashi_Wald}. We adopt Dirichlet boundary conditions. As in~\cite{OrtizBH}, we do this by subtracting the antipodal Wightman function in ESU from the Wightman function. After pulling back to the circular trajectory~\eqref{eqn:: trajectory}, the thermal, derivative-coupled, Wightman function also enjoys the time-translation symmetry of the linearly coupled Wightman function~\eqref{eqn:: Wightman AdS}. 

The thermal Wightman function~\eqref{eqn:: therm exp} and response function split into vacuum and thermal contributions respectively,
\begin{subequations}
    \begin{align}
    \mc{W}_\beta(s,0)&~=~\mc{W}_\infty(s,0)+\Delta\mc{W}_\beta(s,0)\,,\\
    \FF(E,\beta)&~=~\FF_\infty(E)+\Delta\FF_\beta(E)\,.
\end{align}
\end{subequations}
The vacuum contribution $\FF_\infty$ is equal to $E^2$ times the result from Section~\ref{subsec:: large boundary zero}. We now focus on the contribution due to the ambient temperature, $\Delta\FF_\beta$. 

In the long-interaction-time limit, $X\to1$, $\Delta\FF_\beta$ is given by 
\begin{equation}\label{eqn:: thermal response}
    \Delta\FF_\beta(E)~=~\int\dd s\,\ee^{-\ii Es}\Delta\mc{W}_\beta(s,0)\,,
\end{equation}where
    \begin{multline}
        \Delta\mc{W}_\beta(s,0)~=~\frac{\gamma^2}{4\pi\alpha}\frac{1}{\sqrt{1+\frac{R^2}{\alpha^2}}}\sum_{l=0}^\infty\sum_{m=-l}^l(\omega_l-m\Omega)^2\\\times\frac{(l+|m|)!}{(l-|m|)!}(1-(-1)^l)n(\beta\omega_l)\\\times \lrb{\PP_l^{-|m|}\lr{\frac{1}{\sqrt{1+\frac{R^2}{\alpha^2}}}}}^22\cos((\omega_l-m\Omega)\gamma s)\,.
    \end{multline}
$\gamma$ is given by~\eqref{eqn:: gamma}, the factor $(\omega_l-m\Omega)^2$ comes from the derivative coupling, the ratio of factorials is due to Rodrigues' formula for the associated Legendre polynomials~\cite{NIST}, the factor $(1-(-1)^l)$ comes from the Dirichlet boundary conditions, and $n$ is the Bose thermal factor. Due to the Dirichlet boundary conditions, only odd $l$ contribute to $\Delta\mc{W}_\beta$ and we change summation variable to $l=2k+1$, $k\in\NN_0$. Finally, we introduce the Heaviside theta~\cite{NIST} $\Theta(l+\tfrac{1}{2}-|m|)=\Theta(\alpha\omega_l-|m|)$ to extend the summation in $m$ to the integers $\ZZ$ and exchange the summation order.
    \begin{multline}\label{eqn:: thermal wightman contribution}
        \Delta\mc{W}_\beta(s,0)~=~\frac{\gamma^2}{2\pi\alpha}\frac{1}{\sqrt{1+\frac{R^2}{\alpha^2}}}\sum_{m=-\infty}^\infty\sum_{k=0}^\infty(\omega_k-m\Omega)^2\\\times\Theta(\alpha\omega_k-|m|)\frac{(2k+1+|m|)!}{(2k+1-|m|)!}n(\beta\omega_k)\\\times \lrb{\PP_{2k+1}^{-|m|}\lr{\frac{1}{\sqrt{1+\frac{R^2}{\alpha^2}}}}}^22\cos((\omega_k-m\Omega)\gamma s)\,,
    \end{multline}where now $\omega_k=\frac{1}{\alpha}(2k+\frac{3}{2})$. The thermal contribution to the response function is then
    \begin{multline}\label{eqn:thermal response ads}
        \Delta\FF_\beta(E)~=~\frac{1}{\alpha}\frac{E^2}{\sqrt{1+\frac{R^2}{\alpha^2}}}\sum_{m=-\infty}^\infty\sum_{k=0}^\infty \Theta(\alpha\omega_k-|m|)\\\times n(\beta\omega_k)\frac{(2k+1+|m|)!}{(2k+1-|m|)!}\lrb{\PP_{2k+1}^{-|m|}\lr{\frac{1}{\sqrt{1+\frac{R^2}{\alpha^2}}}}}^2\\
        \times\bigg(\delta(E-\gamma(\omega_k-m\Omega))+\delta(E+\gamma(\omega_k-m\Omega))\bigg)\,.
    \end{multline}
The Dirac deltas in \eqref{eqn:thermal response ads} show that $\Delta\FF_\beta(E)$ is a distribution in~$E$.

\subsection{Large-$\alpha$ limit}\label{sec:: thermal ads large alpha}
In this Section, we consider the large-$\alpha$ asymptotic behaviour of the contribution to the response function due to the ambient temperature $\Delta\FF_\beta$~\eqref{eqn:thermal response ads}.

Given the distributional nature of $\Delta\FF_\beta$~\eqref{eqn:thermal response ads}, we introduce the integrated response contribution due to finite temperature,
\begin{equation}\label{eqn:Gform}
    \GG(\alpha)~=~\int_{\RR}\dd E\, E^{-2}\sigma(E)\Delta\FF_\beta(E)\,,
\end{equation}where $\sigma\in C_0^\infty(\RR)$ is a real-valued function of compact support such that $0\notin\supp\{\sigma\}$ and either $\supp\{\sigma\}\subset\RR_{>0}$ or $\supp\{\sigma\}\subset\RR_{<0}$. In Appendix~\ref{app:G asymptotics}, we show that
\begin{align}\label{eqn:G factored}
    &\GG(\alpha)~=~\nonumber\\
    &\int_\RR\dd E
    \,\sigma(E)\Bigg[\frac{1}{2\Gamma}\sum_{m>\frac{|E|}{\Gamma\Omega}}n(\beta\omega_+)J_{|m|}^2(\omega_+R)\nonumber\\
    &\hphantom{\int_{\RR}\dd E\sigma(E)}\!+\frac{1}{2\Gamma}\sum_{m>-\frac{|E|}{\Gamma\Omega}}n(\beta\omega_-)J_{|m|}^2(\omega_-R)\nonumber\\
    &-\frac{R^2}{16\beta\alpha^2}\ln\lr{\frac\alpha R}\delta(|E|-\Gamma\Omega)+O(\alpha^{-2})\Bigg]\,,
\end{align}where $\omega_\pm=m\Omega\mp |E|/\Gamma$. By comparison with~\eqref{eqn:Gform}, we read off the thermal contribution to the response function $\Delta\FF_\beta$ as
\begin{align}
    \Delta\FF_\beta(E)~=~&\frac{E^2}{2\Gamma}\sum_{m>\frac{|E|}{\Gamma\Omega}}n(\beta\omega_+)J_{|m|}^2(\omega_+R)\nonumber\\
    +&\frac{E^2}{2\Gamma}\sum_{m>-\frac{|E|}{\Gamma\Omega}}n(\beta\omega_+)J_{|m|}^2(\omega_+R)\nonumber\\
    -\frac{R^2E^2}{16\beta\alpha^2}&\ln\lr{\frac\alpha R}\delta(|E|-\Omega\Gamma)+O(\alpha^{-2})\,,\label{eqn:ads therm response}
\end{align}where the $O$-notation is used in a distributional sense.

The leading correction to~\eqref{eqn:ads therm response} is order $\alpha^{-2}\ln\alpha$ and the subleading corrections are order $\alpha^{-2}$. We note that the $\alpha^{-2}\ln\alpha$ term has no counterpart in a similar expansion for a field confined within a finite cylinder in Minkowski spacetime, where the leading correction is inversely proportional to the square of the cylinder radius~\cite{BunneyBoundary}.

\subsection{Static observers}
In the static observer limit, $\Omega\to0$, \eqref{eqn:ads therm response} reduces to
    \begin{align}
        \lim_{\Omega\to0}\Delta\FF_\beta(E)&~=~\frac{1}{2}n(\beta|E|)E^2\sum_{m\in\ZZ}J_{|m|}(ER/\Gamma)\,,
        \nonumber
        \\\label{eqn:: F static 0a}
        &~=~\frac{1}{2}E^2n(\beta|E|)\,,
    \end{align}
using Neumann's addition formula~\cite{NIST}, $\sum_{m\in\ZZ}J^2_{|m|}(x)=1$.

\section{De Sitter spacetime}\label{sec:: ds}
In this Section, we consider the spacetime of constant positive curvature, de Sitter spacetime. Specifically, we consider the static patch of de Sitter, to which we refer as Rindler-de Sitter (RdS) spacetime, a static, globally hyperbolic spacetime of physical and cosmological interest~\cite{AkhmedovDS}. The name Rindler-de Sitter is due to the fact that the restriction of the de Sitter vacuum, the Euclidean (or Bunch-Davies~\cite{BunchDavies} or Chernikov-Tagirov~\cite{ChernikovTagirov}) vacuum, to the static patch is thermal~\cite{GibbonsHawkingDeSitter,LapedesdeSitter} in temperature $1/(2\pi\alpha)$, where $\alpha$ is the de Sitter radius --- akin to the restriction of the Minkowski vacuum to the Rindler patch~\cite{Fulling,Davies1975, Unruh}. We first consider a field prepared in a thermal state with arbitrary temperature and then specialise to the Euclidean vacuum.

\subsection{Spacetime, field, and detector preliminaries}
We consider $(2+1)$-dimensional de Sitter spacetime as embedded in $3+1$ Minkowski spacetime $\RR^{3,1}$ with the embedding equation
\begin{equation}\label{eqn:dS embedding equation}
    \eta^{3,1}_{AB}X^AX^B~=~\alpha^2\,,
\end{equation}where $X^A=(X^0,X^1,X^2,X^3)$ are the coordinates on $\RR^{3,1}$, $\eta^{3,1}=\text{diag}(-1,1,1,1)$ and $\alpha>0$ is the de Sitter radius. The cosmological constant $\Lambda$ is related to $\alpha$ by $\Lambda = 1/\alpha^2$. The Ricci scalar of $\mrm{dS}_3$ is $\mc{R}=6/\alpha^2$.

We use coordinates $(t,r,\theta)$ in which the metric on $\mrm{RdS}$ is given by
\begin{equation}\label{eqn:: ds metric}
    \dd s^2~=~-\lr{1-\frac{r^2}{\alpha^2}}\dd t^2+\frac{\dd r^2}{1-\frac{r^2}{\alpha^2}}+r^2\dd\theta^2\,,
\end{equation}where $-\infty<t<\infty$, $0\leq r<\alpha$, and $0\leq\theta<2\pi$, with the usual coordinate singularity at $r=0$ and $\theta$ periodicially identified with period $2\pi$. Note the presence of a cosmological horizon at $r=\alpha$. The embedding in the embedding space $\RR^{3,1}$ is given by
\begin{subequations}
    \begin{align}
        X^0&~=~\sqrt{\alpha^2-r^2}\sinh(t/\alpha)\,,\\
        X^1&~=~\sqrt{\alpha^2-r^2}\cosh(t/\alpha)\,,\\
        X^2&~=~r\cos\theta\,,\\
        X^3&~=~r\sin\theta\,.
    \end{align}
\end{subequations} The small cosmological constant limit $\Lambda\rightarrow0^+$ is $\alpha\rightarrow0$, which recovers Minkowski spacetime $\RR^{2,1}$. This can be seen immediately from the form of the metric~\eqref{eqn:: ds metric}.

We consider a quantised, conformally coupled, massless, real scalar field $\Phi$. The field obeys the Klein-Gordon equation~\eqref{eqn:: KG static} and we look for separable solutions of the form $\phi_{\omega m}=g_{\omega m}(r)\ee^{\ii m\theta-\ii\omega t}$ with $\omega>0$ and $m\in\ZZ$. The radial factor obeys then the equation
\begin{multline}
    0~=~\left(\frac{\omega^2}{1-\frac{r^2}{\alpha^2}}-\frac{m^2}{r^2}-\frac{3}{4\alpha^2}\right)g_{\omega m}(r)\\+\frac{1}{r}\frac{\p}{\p r}\lr{r\lr{1-\frac{r^2}{\alpha^2}}\frac{\p}{\p r}g_{\omega m}(r)}\,.
\end{multline} The positive-frequency modes with respect to $\p_t$ that are regular at $r=0$ are then given by

\begin{subequations}\label{eqn:: ds field modes}
    \begin{align}
        \phi_{\omega m}(t,r,\theta)&=\mathcal{A}_{\omega m}\lr{\frac{r}{\alpha}}^{|m|}\lr{1-\frac{r^2}{\alpha^2}}^{-\frac{1}{2}\ii\alpha\omega}\ee^{\ii m\theta-\ii\omega t}\nonumber
        \\\times\,{_2F_1}\!\!&\left(\vphantom{\tfrac{r^2}{\alpha^2}}\tfrac{1}{2}\lr{|m|+\tfrac{1}{2}-\ii\alpha\omega},\tfrac{1}{2}\lr{|m|+\tfrac{3}{2}-\ii\alpha\omega}\right.;\nonumber\\&\qquad\qquad\qquad\qquad\left.|m|+1;\tfrac{r^2}{\alpha^2}\right)\,,\\
        \nonumber\mc{A}_{\omega m}&~=~\frac{\sqrt{\sinh(\pi\alpha\omega)}}{2\sqrt{2}\pi^{3/2}(|m|)!}\,\\
        \times \Gamma\!\left(\tfrac{1}{2}\!\!\right.&\left.\lr{|m|+\tfrac{1}{2}+\ii\alpha\omega}\right)\Gamma\lr{\tfrac{1}{2}\!\lr{|m|+\tfrac{3}{2}+\ii\alpha\omega}}\,,
    \end{align}
\end{subequations}
where $\omega>0$, $m\in\ZZ$, and ${_2F_1}$ is the hypergeometric function~\cite{NIST}. The field modes~\eqref{eqn:: ds field modes} are normalised with respect to the Klein-Gordon norm~\eqref{eqn:: KG norm} with a Kronecker delta in $m$ and Dirac delta in $\omega$. The normalisation factor can be read off by considering the asymptotic form near $r=\alpha$~\cite{Higuchi_dS} and using~\cite[(5.4.3)]{NIST}. For comparison, the field modes for a scalar field in $1+1$ RdS and $3+1$ RdS are given in~\cite{AkhmedovDS} and~\cite{Higuchi_dS} respectively.

We expand the field in the field modes~\eqref{eqn:: ds field modes} as
\begin{equation}
    \Phi(t,r,\theta)~=~\int \dd\omega\sum_{m\in\ZZ}\phi_{\omega m}(t,r,\theta)a_{\omega m}+\text{H.c.}\,,
\end{equation}
where the annihilation $a_{\omega m}$ and creation $a^\dagger_{\omega m}$ operators obey the commutation relation $[a_{\omega m},a^\dagger_{\omega' m'}]=\delta(\omega-\omega')\delta_{m,m'}$. The Hilbert space is a Fock space with a Fock vacuum $\ket{0}$ satisfying $a_{\omega m}\ket{0}$ for all $\omega$ and $m$. We shall consider the Fock vacuum, a thermal state in a general inverse temperature $\beta$, defined as in~\eqref{eqn:: therm exp}, and the Euclidean vacuum, in which $\beta=2\pi\alpha$.

We probe the field with a pointlike detector in uniform circular motion,
\begin{subequations}\label{eqn:: trajectory dS}
    \begin{align}
    (t,r,\theta)&~=~(\gamma\tau,R,\Omega\gamma\tau)\,,\\
    \label{eqn:: gamma RdS}
    \gamma&~=~\frac{1}{\sqrt{1
-R^2\Omega^2-\frac{R^2}{\alpha^2}}}\,,
\end{align}
\end{subequations} where $\tau$ is proper time, $R>0$ is the value of the radial coordinate, and $\Omega=\tfrac{\dd\theta}{\dd t}>0$ is the angular velocity with respect to time $t$ in~\eqref{eqn:: ds metric}. We assume that the worldline is timelike, $R\Omega<\sqrt{1-R^2/\alpha^2}$. In the large-$\alpha$ limit, we have $\lim_{\alpha\rightarrow\infty}\gamma=(1-R^2\Omega^2)^{-1/2}=:\Gamma$, where $\Gamma$ is the Lorentz factor associated with circular motion in Minkowski spacetime $\RR^{2,1}$. We couple the detector to the field $\Phi$ with the derivative-coupled interaction Hamiltonian~\eqref{eqn:: deriv interact}, as described in Section~\ref{sec:: thermal prelims}, in the long-interaction-time limit $\chi\to1$. 

In the thermal state in inverse temperature $\beta$, the Wightman function that appears in the derivative-coupled detector response in $\mrm{RdS}$, pulled back to the circular trajectory~\eqref{eqn:: trajectory dS}, is given by
\begin{widetext}
\begin{multline}\label{eqn:thermal Wightman dS}
\mc{W}_\beta(s,0)~=~\sum_{m\in\ZZ}\int_0^\infty\dd\omega\,|\mc{A}_{\omega m}|^2\gamma^2(\omega-m\Omega)^2\lr{
    \frac{R}{\alpha}}^{2|m|}\lr{(1+n(\beta\omega))\ee^{\ii(m\Omega-\omega)\gamma s}+n(\beta\omega)\ee^{-\ii(m\Omega-\omega)\gamma s}}
    \\\times\left|{_2F_1}\lr{\tfrac{1}{2}\lrc{|m|+\tfrac{1}{2}-\ii\alpha\omega},\tfrac{1}{2}\lrc{|m|+\tfrac{3}{2}-\ii\alpha\omega};|m|+1;\tfrac{R^2}{\alpha^2}}\right|^2\,,
\end{multline}where $n(x)=(\ee^x-1)^{-1}$ is the Bose thermal factor as in Section~\ref{sec:thermal states}.

The response function~\eqref{eqn:: therm response} (in the long-time limit) is calculated from~\eqref{eqn:thermal Wightman dS} by first integrating over $s$ and then integrating over $\omega$, leading to
\begin{subequations}\label{eqn: dS response split}
\begin{align}
    \FF(E,\beta)&~=~\FF_\infty(E)+\Delta\FF_\beta(E)\,,\\
    \FF_\infty(E)&~=~\frac{2\pi E^2}{\gamma}\sum_{m>E/(\gamma\Omega)}\lr{\frac{R}{\alpha}}^{2|m|}|\mc{A}_{\varpi_mm}|^2\left|{_2F_1}\lr{\tfrac{1}{2}\lrc{|m|+\tfrac{1}{2}-\ii\alpha\varpi_m},\tfrac{1}{2}\lrc{|m|+\tfrac{3}{2}-\ii\alpha\varpi_m};|m|+1;\tfrac{R^2}{\alpha^2}}\right|^2\,,\label{eqn: dS response vacuum split}
\end{align}
\begin{multline}
    \Delta\FF_\beta(E)~=~\\
       \frac{2\pi E^2}{\gamma}\sum_{m>|E|/(\gamma\Omega)}\!\!\!n(\beta\omega_m^+)\lr{\frac{R}{\alpha}}^{2|m|}|\mc{A}_{\omega_m^+m}|^2\left|{_2F_1}\lr{\tfrac{1}{2}\lrc{|m|+\tfrac{1}{2}-\ii\alpha\omega_m^+},\tfrac{1}{2}\lrc{|m|+\tfrac{3}{2}-\ii\alpha\omega_m^+};|m|+1;\tfrac{R^2}{\alpha^2}}\right|^2
        \\+\frac{2\pi E^2}{\gamma}\sum_{m>-|E|/(\gamma\Omega)}\!\!\!n(\beta\omega_m^-)\lr{\frac{R}{\alpha}}^{2|m|}|\mc{A}_{\omega_m^-m}|^2\left|{_2F_1}\lr{\tfrac{1}{2}\lrc{|m|+\tfrac{1}{2}-\ii\alpha\omega_m^-},\tfrac{1}{2}\lrc{|m|+\tfrac{3}{2}-\ii\alpha\omega_m^-};|m|+1;\tfrac{R^2}{\alpha^2}}\right|^2\,,\label{eqn:thermal contribution response dS}
\end{multline}
\end{subequations}where $\varpi_m=m\Omega-E/\gamma$ and $\omega_m^\pm=m\Omega\mp |E|/\gamma$. The response function~\eqref{eqn: dS response split} naturally decomposes into a contribution from the static vacuum $\FF_\infty$ and a contribution due to the ambient temperature $\Delta\FF_\beta$. It is clear that $\Delta\FF_\beta$ is an even function of the energy gap $E$.
\end{widetext}

\subsection{Large-$\alpha$ limit}
In this Section, we consider the large-$\alpha$ limit of the detector response~\eqref{eqn: dS response split} when the field is  prepared in the Euclidean vacuum. When restricted to the static patch, this is a thermal state in inverse temperature $\beta=2\pi\alpha$, in which case the Wightman function is de Sitter invariant~\cite{AkhmedovDS,AllenVacuumStatesdS}.

\subsubsection{Thermal contribution}\label{subsubsec: thermal}
We first consider the thermal contribution $\Delta\FF_\beta$ to the response function~\eqref{eqn: dS response split}. By a simple bounding argument, we demonstrate that $\Delta\FF_\beta$ is exponentially suppressed compared to the static vacuum contribution $\FF_\infty$.

Let $m^\pm=1+\lfloor\pm|E|/(\gamma\Omega)\rfloor$, the first value of $m$ in each summand in~\eqref{eqn:thermal contribution response dS}, where the notation suppresses the $E$-dependence of $m^\pm$. It follows that
\begin{equation}\label{eqn: omega inequality}
    0~<~\omega_{m^\pm}^\pm~=~m^\pm\Omega\mp|E|/\gamma~\leq ~\Omega\,.
\end{equation}As $n(x)$ is decreasing in $x $, we have $n(\beta\omega_m^\pm)\leq n(\beta\omega_{m^\pm}^\pm)$. Each of the two sums in $\Delta\FF_\beta$~\eqref{eqn:thermal contribution response dS} can hence be bounded above by a replacing $n(\beta\omega^\pm_m)$ by $n(\beta\omega^\pm_{m^\pm})$, giving the bound
\begin{equation}\label{eqn:Fbeta inequal}
    \Delta\FF_\beta(E)\leq n(\beta\omega_{m^+}^+)\FF_\infty(|E|)+n(\beta\omega_{m^-}^-)\FF_\infty(-|E|)\,,
\end{equation}where we have used~\eqref{eqn: dS response vacuum split}. 

As $\beta=2\pi\alpha$, the thermal factors $n(\beta\omega_{m^\pm}^\pm)\sim\exp(-2\pi\alpha\omega^\pm_{m^\pm})$ in~\eqref{eqn:Fbeta inequal} are exponentially suppressed for large $\alpha$, as follows from~\eqref{eqn: omega inequality}. The static vacuum factors $\FF_\infty$ in~\eqref{eqn:Fbeta inequal} have the large-$\alpha$ limit given by
\begin{equation}
    \lim_{\alpha\to\infty}\FF_\infty(E)~=~\frac{E^2}{2\Gamma}\sum_{m>E/(\Omega\Gamma)}J_{|m|}^2(mv-ER/\Gamma)\,,
\end{equation}which is the response function for a detector undergoing uniform circular motion in $2+1$ Minkowski spacetime~\cite{Upton}. Hence, $\Delta\FF_\beta$ is exponentially suppressed in the large-$\alpha$ limit. Note, however, that the coefficient of $\alpha$ in the exponent is not continuous in $E$ and the parameters of the motion, and this coefficient can take arbitrarily small positive values.

\subsubsection{Full Euclidean vacuum}
We now consider the full response in the Euclidean vacuum. The Wightman function is expressible in terms of the hypergeometric function, derivable by exploiting the symmetries of de Sitter spacetime~\cite{AllenVacuumStatesdS,MoschelladSGeneralWightman}. In our case of a conformally coupled, massless scalar field, the Wightman function is
\begin{subequations}\label{eqn:vacuum Wightman}
\begin{align}
    \mc{W}_{\mrm{Euc}}(\xx,\xx')&~=~\frac{1}{4\pi}\frac{1}{\sqrt{\sigma(\xx,\xx')}}\,,\\
    \sigma(\xx,\xx')&~=~\eta^{3,1}_{AB}\lr{X^A-{X'}^A}\lr{X^B-{X'}^B}\,,
\end{align}    
\end{subequations}where $\sigma$ is the geodesic squared distance in the embedding space and we momentarily suppress the distributional character.

Pulled back to the circular trajectory~\eqref{eqn:: trajectory dS}, the Wightman function $\WW_{\mrm{Euc}}$ is again time-translation invariant. The geodesic squared distance is given by
\begin{multline}\label{eqn:: ds geodesic squared}
    \sigma(s,0)~=~-4\lr{\alpha^2-R^2}\sinh^2\lr{\frac{\gamma s}{2\alpha}}\\+4R^2\sin^2\lr{\frac{\Omega \gamma s}{2}}\,,
\end{multline} and the pullback of $\WW_{\mrm{Euc}}$ to the circular trajectory is
\begin{equation}
    \WW_{\mrm{Euc}}(s,0)~=~\frac{1}{4\pi}\frac{1}{\sqrt{\sigma(s-\ii\varepsilon,0)}}\,,
\end{equation}
where the distributional character has been restored and encoded as the limit $\varepsilon\to0^+$. 
As in the Minkowski vacuum, the square root in the denominator is positive imaginary for positive $s$ and negative imaginary for negative $s$ \cite{HowOftenClick,Biermann}. 

The only distributional contribution to the response function $\FF_{\mrm{Euc}}$ is at $s=0$. Isolating this, we find~\cite{HowOftenClick,Biermann}
\begin{align}\label{eqn:euclidean response}
    \FF_{\mrm{Euc}}(E)&~=~\frac{E^2}{4\pi}\int_{\RR}\dd s\,\frac{\ee^{-\ii Es}}{\sqrt{\sigma(s-\ii\varepsilon,0)}}\nonumber\\
    &~=~\frac{E^2}{4}-\frac{E^2}{2\pi}\int_0^\infty\dd s\,\frac{\sin(Es)}{\sqrt{-\sigma(s,0)}}\,.
\end{align}

It is convenient to parametrise the circular trajectory in terms of $R$ and $V$, where
\begin{equation}
    V~\coloneq~\frac{R\Omega}{\sqrt{1-\frac{R^2}{\alpha^2}}}\,.
\end{equation}Geometrically, $V$ is the velocity of the detector as seen by a static observer at $r=R$. Changing the integration variable to $z=\gamma s(2\alpha)^{-1}$, $\FF_{\mrm{Euc}}$~\eqref{eqn:euclidean response} reads
    \begin{align}\label{eqn:vacuum new params}
        \FF_{\mrm{Euc}}(E)&~=~\frac{E^2}{4}-\frac{E^2}{2\pi\Gamma_V}\int_0^\infty\dd z\,\frac{\sin\lr{\frac{2ER}{\Gamma_VV}z}}{z}\nonumber\\&\times\frac{1}{\sqrt{\sinh^2(\eta z)/(\eta z)^2-V^2\sin^2(z)/z^2}}\,,
    \end{align}where we have written $\Gamma_V\coloneq(1-V^2)^{-1/2}$ and $\eta\coloneq(\Omega\alpha)^{-1}=V^{-1}(\alpha^2/R^2-1)^{-1/2}$.

We show in Appendix~\ref{app:integral expand} that the asymptotic behaviour of~\eqref{eqn:vacuum new params} as $\eta\to0$ with $E$, $R$, and $V$ fixed, is given by
\begin{multline}\label{eqn:dS vacuum expansion}
    \FF_{\mrm{Euc}}(E)~=~\frac{E^2}{4}\\-\frac{E^2}{2\pi\Gamma_V}\int_0^\infty\dd z\,\frac{\sin\lr{\frac{2ER}{\Gamma_VV}z}}{z}\frac{1}{\sqrt{1-V^2\sin(z)^2/z^2}}\\\phantom{ad}+\frac{E^2}{12\pi\Gamma_V}\eta^2\int_0^\infty \dd z\,z\sin\lr{\frac{2ER}{\Gamma_VV}z}\hfill\\\times\bigg(\frac{1}{(1-V^2\sin^2(z)/z^2)^{3/2}}-1\bigg)+o(\eta^2)\,,
\end{multline}where $\eta^2=V^{-2}(\alpha^2/R^2-1)^{-1}=R^2(V\alpha)^{-2}+o(\alpha^{-2})$ in terms of the de Sitter radius of curvature as $\alpha\to\infty$. 

The leading term in~\eqref{eqn:dS vacuum expansion} is the response of a detector undergoing uniform circular motion in $(2+1)$-dimensional Minkowski spacetime with orbital speed $V$ and angular velocity $V/R$~\cite{Biermann}. The next-to-leading term in~\eqref{eqn:dS vacuum expansion} is proportional to $\alpha^{-2}$. As this term is larger than the exponentially small thermal corrections found in Section~\ref{subsubsec: thermal}, we may interpret this term as a curvature correction.

\section{Conclusions}\label{sec:: conclusions}
We addressed the effects of thermality and finite size on an Unruh-DeWitt detector undergoing circular motion in $(2+1)$-dimensional de Sitter and anti-de Sitter spacetimes. In AdS, in which the negative cosmological constant $\Lambda=-\alpha^{-2}$ provides a notion of finite size, we considered a field prepared in the global static vacuum and in a thermal state. In the near-Minkowski limit, $\alpha\to\infty$, we found the leading and subleading asymptotic behaviour of the response function, interpreting it in terms of the response of a detector undergoing circular motion in Minkowski spacetime with a cylindrical boundary. We found that the resonance peaks in the detector's response closely match those of a detector in Minkowski spacetime but with curvature corrections. For an initial thermal state, however, we found a resonance peak with no corresponding term in Minkowski spacetime with a cylindrical boundary.

In RdS, in which the positive cosmological constant $\Lambda=\alpha^{-2}$ provides a notion of ambient temperature, we considered a field prepared in the Euclidean vacuum. The restriction of the Euclidean vacuum to the static patch is a thermal state in temperature $T=(2\pi\alpha)^{-1}$, and the near-Minkowski limit, $\alpha\to\infty$, corresponds to the low-temperature limit $T\to0$. We found that the detector response was dominated by the contribution due to the static vacuum, whilst the temperature corrections decayed exponentially as $\Lambda\to0$. The leading curvature correction to the detector's response is proportional to $\Lambda$.

Our physical motivation to consider RdS and AdS in the limits of small cosmological constants was that their respective positive and negative curvatures provide notions of an ambient temperature and spatial confinement respectively, both of which notions will feature in flat analogue spacetime realisations of the (circular motion) Unruh effect. A curved analogue spacetime realisation of the Unruh effect in RdS and AdS spacetimes may be opened up by recent experiments in a two-dimensional Bose-Einstein condensate, which have realised cosmological spacetimes with positive and negative spatial curvature~\cite{OberthalerDeSitter}.

\section*{Acknowledgements}

We thank Joel Feinstein for discussions on asymptotics of integrals. 
The work of JL was supported by United Kingdom Research and Innovation Science and Technology Facilities Council [grant numbers ST/S002227/1, ST/T006900/1 and ST/Y004523/1].
For the purpose of open access, the authors have applied a CC BY public copyright licence to any Author Accepted Manuscript version arising.

\appendix
\onecolumngrid

\section{Large-$\alpha$ asymptotics in CAdS}\label{app:G asymptotics}
In this Appendix, we demonstrate the asymptotic behaviour of the integrated response contribution due to finite temperature $\GG$~\eqref{eqn:Gform} in the large-$\alpha$ regime leading to the results of~\ref{sec:: thermal ads large alpha}.

\subsection{Decomposition of integrated response due to finite temperature}
The integrated response contribution due to finite temperature $\GG$~\eqref{eqn:Gform} with $\Delta\FF_\beta$~\eqref{eqn:thermal response ads} reads
\begin{multline}\label{eqn:G alpha}
    \GG(\alpha)~=~\frac{1}{\alpha}\frac{1}{\sqrt{1+\frac{R^2}{\alpha^2}}}\sum_{m=-\infty}^\infty\sum_{k=0}^\infty \Theta(\alpha\omega_k-|m|) n(\beta\omega_k)\frac{(2k+1+|m|)!}{(2k+1-|m|)!}\lrb{\PP_{2k+1}^{-|m|}\lr{\frac{1}{\sqrt{1+\frac{R^2}{\alpha^2}}}}}^2\\
    \times \lrb{\sigma(\gamma(\omega_k-m\Omega))+\sigma(-\gamma(\omega_k-m\Omega))}\,,
\end{multline}where $\omega_k=(2k+3/2)/\alpha$. 

We recall that $\sigma\in C^\infty_0$. Let $\sigma_I=\inf\supp\{\sigma\}$ and $\sigma_S=\sup\supp\{\sigma\}$. We assume that the support of $\sigma$ is chosen such that either $\supp\{\sigma\}\subset\RR_{>0}$ with $0<\sigma_I<\Omega\Gamma<\sigma_S$ or $\supp\{\sigma\}\subset\RR_{<0}$ with $\sigma_I<-\Omega\Gamma<\sigma_S<0$. Since $\Delta\FF_\beta$ is even in $E$, we shall for now assume that $\supp\{\sigma\}\subset\RR_{>0}$ and relax this assumption in~\eqref{eqn:G final assume relax}.

We rewrite the integrated response contribution due to finite temperature to separate the two $\sigma$ in~\eqref{eqn:G alpha},
\begin{subequations}\label{eqn:Gpm split}
    \begin{align}
       \GG(\alpha)&~=~\GG^+(\alpha)+\GG^-(\alpha)\,,\\ 
       \GG^+(\alpha)&~=~\frac{1}{\alpha}\frac{1}{\sqrt{1+\frac{R^2}{\alpha^2}}}\sum_{m=-\infty}^\infty\sum_{k=0}^\infty \Theta(\alpha\omega_k-|m|) n(\beta\omega_k)\frac{(2k+1+|m|)!}{(2k+1-|m|)!}\lrb{\PP_{2k+1}^{-|m|}\lr{\frac{1}{\sqrt{1+\frac{R^2}{\alpha^2}}}}}^2\label{eqn:Gplus}
    \!\! \sigma(\gamma(\omega_k-m\Omega))\,,\\
    \GG^-(\alpha)&~=~\frac{1}{\alpha}\frac{1}{\sqrt{1+\frac{R^2}{\alpha^2}}}\sum_{m=-\infty}^\infty\sum_{k=0}^\infty \Theta(\alpha\omega_k-|m|) n(\beta\omega_k)\frac{(2k+1+|m|)!}{(2k+1-|m|)!}\lrb{\PP_{2k+1}^{-|m|}\lr{\frac{1}{\sqrt{1+\frac{R^2}{\alpha^2}}}}}^2
    \!\! \sigma(-\gamma(\omega_k-m\Omega))\,.\label{eqn:Gminus}
    \end{align}
\end{subequations}

We recall that $\supp\{\sigma\}\subset[\sigma_I,\sigma_S]$. 
For fixed $m$ in \eqref{eqn:Gplus} and~\eqref{eqn:Gminus}, the values of $k$ in the sums over $k$ are hence restricted by 
\begin{subequations}
\label{eq:k-ineqs-Gplusminus}
    \begin{align}\label{eq:k-ineqs-Gplus}
        \frac{\alpha}{2}\lr{\frac{\sigma_I}{\gamma}+m\Omega}-\frac{3}{4}&~<~k~<~\frac{\alpha}{2}\lr{\frac{\sigma_S}{\gamma}+m\Omega}-\frac{3}{4}\,,\quad\quad \,\,\hphantom{!}\text{in~}\GG^+\,,\\\label{eq:k-ineqs-Gminus}
        \frac{\alpha}{2}\lr{-\frac{\sigma_S}{\gamma}+m\Omega}-\frac{3}{4}&~<~k~<~\frac{\alpha}{2}\lr{-\frac{\sigma_I}{\gamma}+m\Omega}-\frac{3}{4}\,,\quad\quad \text{in~}\GG^-\,.
    \end{align}
\end{subequations}Depending on the value of $m$, the conditions \eqref{eq:k-ineqs-Gplusminus} fall into three different cases. 
For sufficiently large $\alpha$, with the other parameters fixed, these cases are as follows. 

In $\GG^+$:
    \begin{itemize}
        \item[-] For $m \leq -\frac{\sigma_S}{\gamma\Omega}$, no $k$ satisfy~\eqref{eq:k-ineqs-Gplus}.
        \item[-] For $-\frac{\sigma_S}{\gamma\Omega}<m\leq-\frac{\sigma_I}{\gamma\Omega}$, $k$ satisfies $0\leq k\leq\Kmax^+ \coloneq \frac{\alpha}{2}(\frac{\sigma_S}{\gamma}+m\Omega)-\frac{3}{4}$. 
        We denote the set of these $m$ by $\mc{C}^{1+}$. 
        \item[-] For $-\frac{\sigma_I}{\gamma\Omega} < m$, $k$ satisfies $\Kmin^+\leq k\leq\Kmax^+$, where $\Kmin^+ \coloneq \frac \alpha 2(\frac{\sigma_I}{\gamma}+m\Omega)-\frac34$. 
        We denote the set of these $m$ by $\mc{C}^{2+}$. 
    \end{itemize}
In $\GG^-$:
    \begin{itemize}
        \item[-] For $m\leq \frac{\sigma_I}{\gamma\Omega}$, no $k$ satisfy~\eqref{eq:k-ineqs-Gminus}.
        \item[-] For $\frac{\sigma_I}{\gamma\Omega}<m\leq\frac{\sigma_S}{\gamma\Omega}$, $k$ satisfies $0\leq k\leq\Kmax^-\coloneq\frac\alpha 2(-\frac{\sigma_I}{\gamma}+m\Omega)-\frac34$. 
        We denote the set of these $m$ by $\mc{C}^{1-}$. 
        \item[-] For $\frac{\sigma_S}{\gamma\Omega}<m$, $k$ satisfies $\Kmin^-\leq k\leq\Kmax^-$, where $\Kmin^-\coloneq\frac\alpha2(-\frac{\sigma_S}{\gamma}+m\Omega)-\frac34$.
        We denote the set of these $m$ by $\mc{C}^{2-}$. 
    \end{itemize}
We note that the sets $\mc{C}^{1\pm}$ are finite and $\pm1\in\mc{C}^{1\mp}$. Note also that the notation suppresses the dependence of $\Kmax^\pm$ and $\Kmin^\pm$ on $\alpha$ and~$m$. 

With this notation, we may split $\GG^\pm$ as 
\begin{subequations}\label{eqn:G split12}
    \begin{align}
        \GG^\pm(\alpha)&=\GG^{1\pm}(\alpha)+\GG^{2\pm}(\alpha)\,,\\
        \GG^{1\pm}(\alpha)&=\frac{1}{\alpha}\frac1{\sqrt{1+\frac{R^2}{\alpha^2}}}\sum_{m\in\mc{C}^{1\pm}}\sum_{k=0}^{\Kmax^\pm}\!\!\Theta(\alpha\omega_k-|m|)n(\beta\omega_k)\frac{(2k+1+|m|)!}{(2k+1-|m|)!}\lrb{\PP^{-|m|}_{2k+1}\lr{\frac{1}{\sqrt{1+\frac{R^2}{\alpha^2}}}}}^2
       \!\!\!\! \sigma(\pm\gamma(\omega_k-m\Omega))\,,
       \label{eqn:G split12-1}\\
        \GG^{2\pm}(\alpha)&=\frac{1}{\alpha}\frac1{\sqrt{1+\frac{R^2}{\alpha^2}}}\sum_{m\in\mc{C}^{2\pm}}\sum_{k=\Kmin^\pm}^{\Kmax^\pm}\!\!\!\!\!\Theta(\alpha\omega_k-|m|)n(\beta\omega_k)\frac{(2k+1+|m|)!}{(2k+1-|m|)!}\lrb{\PP^{-|m|}_{2k+1}\lr{\frac{1}{\sqrt{1+\frac{R^2}{\alpha^2}}}}}^2
        \!\!\!\!\sigma(\pm\gamma(\omega_k-m\Omega))\,.
        \label{eqn:G split12-2}
    \end{align}
\end{subequations}

\subsection{$\GG^{1\pm}$}
We consider first $\GG^{1\pm}$ \eqref{eqn:G split12-1}. 
To further decompose the sum over $k$, we fix a constant $p \in (0,\frac13)$ and set $K: = \left\lfloor\lr{\frac\alpha R}^p\right\rfloor$, where $\lfloor\cdot\rfloor$ is the floor function~\cite{NIST}.
For sufficiently large $\alpha$, we then have $K < \Kmax^\pm$, and we may write 
\begin{subequations}
    \begin{align}
       \GG^{1\pm}(\alpha)&=\GG^{1\pm}_<(\alpha)+\GG^{1\pm}_>(\alpha)\,,\\ 
       \GG^{1\pm}_<(\alpha)&=\frac{1}{\alpha}\frac1{\sqrt{1+\frac{R^2}{\alpha^2}}}\sum_{m\in\mc{C}^{1\pm}}\sum_{k=0}^{K-1}\!\!\Theta(\alpha\omega_k-|m|)n(\beta\omega_k)\frac{(2k+1+|m|)!}{(2k+1-|m|)!}\lrb{\PP^{-|m|}_{2k+1}\lr{\frac{1}{\sqrt{1+\frac{R^2}{\alpha^2}}}}}^2
       \!\!\!\! \sigma(\pm\gamma(\omega_k-m\Omega))\,,
       \label{eqn:G split12-1-less}\\
       \GG^{1\pm}_>(\alpha)&=\frac{1}{\alpha}\frac1{\sqrt{1+\frac{R^2}{\alpha^2}}}\sum_{m\in\mc{C}^{1\pm}}\sum_{k=K}^{\Kmax^\pm}\!\!\Theta(\alpha\omega_k-|m|)n(\beta\omega_k)\frac{(2k+1+|m|)!}{(2k+1-|m|)!}\lrb{\PP^{-|m|}_{2k+1}\lr{\frac{1}{\sqrt{1+\frac{R^2}{\alpha^2}}}}}^2
       \!\!\!\! \sigma(\pm\gamma(\omega_k-m\Omega))\,.
       \label{eqn:G split12-1-more}
    \end{align}
\end{subequations}

To address $\GG^{1\pm}_<$ \eqref{eqn:G split12-1-less}, 
we observe that the hypergeometric representation of the associated Legendre function \cite[(14.3.1)]{NIST} gives 
\begin{equation}
    \PP^{-|m|}_l(1-\delta) = \frac{1}{\Gamma(1+|m|)}\lr{\frac{\delta}{2}}^{|m|/2}\lr{1+O(\delta l^2)}\,
\end{equation}
as $\delta\to0^+$, with $m$ fixed and $l$ bounded relative to $\delta$ so that $\delta l^2 \to0$. 
In \eqref{eqn:G split12-1-less}, 
$\delta= \frac12 (R/\alpha)^2 \left(1 + O(R^2/\alpha^2) \right)$ and 
$l = 2k+1$, so that $\delta l^2 = O \! \left({k^2/\alpha^2}\right)$, 
and $m$ takes values in the finite set $\mc{C}^{1\pm}$ that is independent of $\alpha$ and contains $\mp1$. 
Elementary estimates, using $0<p< \frac13$ and the Laurent expansion 
\begin{align}
n(x) ~=~ \frac{1}{x} -\frac12 + O(x)
\,, 
\end{align}
show that the contributions from the terms with $|m|>1$ are $o(\alpha^{-2})$, and combining this with similar estimates for the $m=-1$ term in $\GG^{1+}_<$ and the $m=1$ term in $\GG^{1-}_<$ gives
    \begin{align}\nonumber
        \GG_<^{1\pm}(\alpha)&~=~\frac{R^2}{4\alpha^2\beta}\sum_{k=0}^{K-1}\frac{(2k+1)(2k+2)}{(2k+\tfrac{3}{2})}\sigma(\Omega\Gamma)
        + o(\alpha^{-2})\\
    &~=~\frac{R^2}{4\alpha^2\beta}\sigma(\Omega\Gamma)\lr{K^2+\frac{1}{2}K-\frac{1}{8}\psi^{(0)}\lr{K+\frac{3}{4}}-\frac{1}{6}+\frac{1}{8}\psi^{(0)}\lr{\frac{7}{4}}}+o(\alpha^{-2})
    \nonumber\\
    &~=~\frac{R^2}{4\alpha^2\beta}\sigma(\Omega\Gamma)\lr{K^2+\frac{1}{2}K-\frac{1}{8}\ln\lr{K+\frac{3}{4}}-\frac{1}{6}+\frac{1}{8}\psi^{(0)}\lr{\frac{7}{4}}}+o(\alpha^{-2})\,, 
    \label{eqn:Gpm1<}
    \end{align}
where in the first equality we have evaluated the sum over $k$ in terms of the digamma function $\psi^{(0)}$ and in the second equality used the asymptotic formula 
$\psi^{(0)}(z) = \ln(z)+ O(1/z)$ as $z\to\infty$~\cite{NIST}. 

To address $\GG^{1\pm}_>$~\eqref{eqn:G split12-1-more}, we observe that Stirling's approximation~\cite{NIST} gives
\begin{equation}\label{eqn:Stirling expand}
    \frac{(2k+1+|m|)!}{(2k+1-|m|)!}~=~\lr{2k+\frac{3}{2}}^{2|m|}\lr{1-\frac{|m|(4|m|^2-1)}{12\lr{2k+\frac{3}{2}}^2}+O(k^{-4})}\,,
\end{equation}as $k\to\infty$, with $m$ fixed.

The asymptotic behaviour of the associated Legendre function $\PP_l^{-|m|}(\cos\varphi)$ uniformly in $m$ in the double limit $l\to\infty$ and $\varphi\to0$ is given by~\cite{MagnusOberhettinger}~\cite[(8.722)]{G+R}. Application of this to the associated Legendre function in $\GG^{1\pm}_>$~\eqref{eqn:G split12-1-more} gives
\begin{multline}\label{eqn:P expand}
    \PP_{2k+1}^{-|m|}\lr{\frac{1}{\sqrt{1+\frac{R^2}{\alpha^2}}}}~=~\lr{2k+\frac{3}{2}}^{-|m|}\lr{1-\frac{R^2}{8\alpha^2}}^{-|m|}\\\times\bigg(J_{|m|}(\omega_kR)+\frac{R^2}{4\alpha^2}\lr{\frac{J_{|m|+1}(\omega_kR)}{2\omega_kR}-J_{|m|+2}(\omega_kR)+\frac{\omega_kR}{6}J_{|m|+3}(\omega_kR)}\bigg)+O(\alpha^{-4})\,.
\end{multline}

Elementary estimates, using $0<p<\tfrac13$, the expansions~\eqref{eqn:Stirling expand} and~\eqref{eqn:P expand}, and the Euler-Maclaurin formula
\begin{multline}\label{eqn:euler-mac}
    \sum_{n=p}^qg(n)~=~\int_p^q\dd x\,g(x)+\frac{1}{2}(g(p)+g(q))+\sum_{i=2}^l\frac{b_i}{i!}\left(g^{(i-1)}(q)-g^{(i-1)}(p)\right)+\int_p^q  \dd x\,\frac{\widetilde{B}_l(1-x)}{l!}g^{(l)}(x)\,,
\end{multline}where $l$ is any positive integer and $\widetilde{B}_j$ and $b_j$ are the periodic Bernoulli polynomials and Bernoulli numbers~\cite{NIST}, give
\begin{align}\label{eqn:Gpm1>}
    \GG_>^{1\pm}(\alpha)~=~&\sum_{m\in\mc{C}^{1\pm}}\frac{1}{2}\int_0^\infty\dd\omega\,n(\beta\omega)J_{|m|}^2(\omega R)\sigma(\pm\Gamma(\omega-m\Omega))\nonumber\,\\
    -&\frac{R^2\sigma(\Omega\Gamma)}{4\beta\alpha^2}\lr{K^2+\frac{1}{2}K+\frac{103}{12}-\frac{1}{8}\ln(2)}+\frac{R^2}{32\beta\alpha^2}\ln(K+\tfrac{3}{4})\sigma(\Omega\Gamma)\nonumber\\
    -&\frac{R^2}{32\beta\alpha^2}\ln\lr{\frac{\alpha}{R}}\sigma(\Omega\Gamma)\nonumber\\
    \mp&\frac{R^2\Gamma^3}{4\alpha^2}\sum_{m\in\mc{C}^{1\pm}}\int_0^\infty\dd\omega\,\td{}{\omega}\lrb{n(\beta\omega)(\omega-m\Omega)J_{|m|}^2(\omega R)}\sigma(\pm\Gamma(\omega-m\Omega))\nonumber\\
    -&\frac{1}{24\alpha^2}\sum_{m\in\mc{C}^{1\pm}\setminus\{\mp1\}}|m|(4|m|^2-1)\int_0^\infty\dd\omega\,\frac{n(\beta\omega)}{\omega^2}J_{|m|}^2(\omega R)\sigma(\pm\Gamma(\omega-m\Omega))\nonumber\\
    -&\frac{1}{8\alpha^2}\int_0^\infty\dd\omega\,\bigg\{\frac{n(\beta\omega)}{\omega^2}J^2_{1}(\omega R)\sigma(\pm\Gamma(\omega\pm\Omega))-\frac{R^2}{4\beta}\frac{1}{\omega}\sigma(\Omega\Gamma)\Theta\lr{\frac{1}{R}-\omega}\bigg\}\nonumber\\
    +&\frac{R^2}{4\alpha^2}\sum_{m\in\mc{C}^{1\pm}}\int_0^\infty\dd\omega\,n(\beta\omega)J_{|m|}(\omega R)\bigg\{\bigg(\frac{|m|}{2}-1\bigg)J_{|m|}(\omega R)+\frac{1}{2\omega R}J_{|m|+1}(\omega R)\nonumber\\-&J_{|m|+2}(\omega R)+\frac{\omega R}{6}J_{|m|+3}(\omega R)\bigg\}\sigma(\pm\Gamma(\omega-m\Omega))+o(\alpha^{-2})\,.
\end{align}

Adding $\GG^{1\pm}_<$~\eqref{eqn:Gpm1<} and $\GG^{1\pm}_>$~\eqref{eqn:Gpm1>}, we find
\begin{align}\label{eqn:G1pm first expand}
    \GG^{1\pm}(\alpha)~=~&\sum_{m\in\mc{C}^{1\pm}}\frac{1}{2}\int_0^\infty\dd\omega\,n(\beta\omega)J_{|m|}^2(\omega R)\sigma(\pm\Gamma(\omega-m\Omega))\nonumber\,\\
    -&\frac{R^2}{32\beta\alpha^2}\ln\lr{\frac{\alpha}{R}}\sigma(\Omega\Gamma)
    -\frac{R^2\sigma(\Omega\Gamma)}{4\beta\alpha^2}\lr{\frac{35}{4}-\frac{1}{8}\ln(2)-\frac{1}{8}\psi^{(0)}\lr{\frac{7}{4}}}\nonumber\\
    \mp&\frac{R^2\Gamma^3}{4\alpha^2}\sum_{m\in\mc{C}^{1\pm}}\int_0^\infty\dd\omega\,\td{}{\omega}\lrb{n(\beta\omega)(\omega-m\Omega)J_{|m|}^2(\omega R)}\sigma(\pm\Gamma(\omega-m\Omega))\nonumber\\
    -&\frac{1}{24\alpha^2}\sum_{m\in\mc{C}^{1\pm}\setminus\{\mp1\}}|m|(4|m|^2-1)\int_0^\infty\dd\omega\,\frac{n(\beta\omega)}{\omega^2}J_{|m|}^2(\omega R)\sigma(\pm\Gamma(\omega-m\Omega))\nonumber\\
    -&\frac{1}{8\alpha^2}\int_0^\infty\dd\omega\,\bigg\{\frac{n(\beta\omega)}{\omega^2}J^2_{1}(\omega R)\sigma(\pm\Gamma(\omega\pm\Omega))-\frac{R^2}{4\beta}\frac{1}{\omega}\sigma(\Omega\Gamma)\Theta\lr{\frac{1}{R}-\omega}\bigg\}\nonumber\\
    +&\frac{R^2}{4\alpha^2}\sum_{m\in\mc{C}^{1\pm}}\int_0^\infty\dd\omega\,n(\beta\omega)J_{|m|}(\omega R)\bigg\{\bigg(\frac{|m|}{2}-1\bigg)J_{|m|}(\omega R)+\frac{1}{2\omega R}J_{|m|+1}(\omega R)\nonumber\\
    -&J_{|m|+2}(\omega R)+\frac{\omega R}{6}J_{|m|+3}(\omega R)\bigg\}\sigma(\pm\Gamma(\omega-m\Omega))+o(\alpha^{-2})\,.
\end{align}Note that whilst $\GG_<^{1\pm}$~\eqref{eqn:Gpm1<} and $\GG_>^{1\pm}$ \eqref{eqn:Gpm1>} individually depend on the auxiliary function~$K$, 
the $K$-dependence has cancelled out in $\GG^{1\pm}$~\eqref{eqn:G1pm first expand}. 

The leading and subleading terms in~\eqref{eqn:G1pm first expand} are given by
\begin{equation}\label{eqn:G1expansion}
    \GG^{1\pm}(\alpha)~=~\sum_{m\in\mc{C}^{1\pm}}\frac{1}{2}\int_0^\infty\dd\omega\,n(\beta\omega)J_{|m|}^2(\omega R)\sigma(\pm\Gamma(\omega-m\Omega))
    -\frac{R^2}{32\beta\alpha^2}\ln\lr{\frac{\alpha}{R}}\sigma(\Omega\Gamma)+O(\alpha^{-2})\,.
\end{equation}

\subsection{$\GG^{2\pm}$}
We consider next $\GG^{2\pm}$~\eqref{eqn:G split12}. To decompose the sum over $m$, we fix a constant $q\in(0,\tfrac12)$ and set $M\coloneq\left\lfloor\lr{\frac{\alpha}{R}}^q\right\rfloor$. For sufficiently large $\alpha$, the argument of the Heaviside theta is always positive and we may write
\begin{subequations}\label{eqn:G2 split}
    \begin{align}
        \GG^{2\pm}(\alpha)&~=~\GG^{2\pm}_<(\alpha)+\GG^{2\pm}_>(\alpha)\,,\\
        \GG^{2\pm}_<(\alpha)&~=~\frac{1}{\alpha}\frac1{\sqrt{1+\frac{R^2}{\alpha^2}}}\sum_{\substack{m\in\mc{C}^{2\pm}\!\!\!\!\!,\\ m\leq M-1}}\sum_{k=\Kmin^\pm}^{\Kmax^\pm}\!\!\!\!\!n(\beta\omega_k)\frac{(2k+1+|m|)!}{(2k+1-|m|)!}\lrb{\PP^{-|m|}_{2k+1}\lr{\frac{1}{\sqrt{1+\frac{R^2}{\alpha^2}}}}}^2
        \!\!\!\!\sigma(\pm\gamma(\omega_k-m\Omega))\,,\label{eqn:G2split msmall}\\
        \GG^{2\pm}_>(\alpha)&~=~\frac{1}{\alpha}\frac1{\sqrt{1+\frac{R^2}{\alpha^2}}}\sum_{m=M}^\infty\sum_{k=\Kmin^\pm}^{\Kmax^\pm}\!\!\!\!\!n(\beta\omega_k)\frac{(2k+1+|m|)!}{(2k+1-|m|)!}\lrb{\PP^{-|m|}_{2k+1}\lr{\frac{1}{\sqrt{1+\frac{R^2}{\alpha^2}}}}}^2
        \!\!\!\!\sigma(\pm\gamma(\omega_k-m\Omega))\,.\label{eqn:G2split mlarge}
    \end{align}
\end{subequations}

To address $\GG^{2\pm}_>$~\eqref{eqn:G2split mlarge}, we observe that the connection formula for the associated Legendre function~\cite[(14.9.13)]{NIST} gives
\begin{equation}\label{eqn:Legendre connection}
    \frac{(2k+1+|m|)!}{(2k+1-|m|)!}\lrb{\PP^{-|m|}_{2k+1}\lr{\frac{1}{\sqrt{1+\frac{R^2}{\alpha^2}}}}}^2~=~\frac{(2k+1-|m|)!}{(2k+1+|m|)!}\lrb{\PP^{+|m|}_{2k+1}\lr{\frac{1}{\sqrt{1+\frac{R^2}{\alpha^2}}}}}^2\,.
\end{equation}Application of the connection formula~\eqref{eqn:Legendre connection} and upper bound~\cite[(6)]{LohferAssociatedLegendreInequal} gives
\begin{equation}\label{eqn:P inequal}
    \frac{(2k+1-|m|)!}{(2k+1+|m|)!}\lrb{\PP^{+|m|}_{2k+1}\lr{\frac{1}{\sqrt{1+\frac{R^2}{\alpha^2}}}}}^2~<~\frac{\Gamma^2\lr{\frac14}\ee^{1/2}}{\pi^2}\,.
\end{equation}Elementary estimates, using~\eqref{eqn:P inequal}, show
    \begin{equation}\label{eqn:G2max inequal}
        \left|\GG^{2\pm}_>(\alpha)\right|~<~\frac{A}{2\beta}\frac{\ee^{-\beta\Omega M}}{1-\ee^{-\beta\Omega}}~=~O(\ee^{-\beta\Omega\lfloor(\alpha /R)^q\rfloor})\,,
    \end{equation}where $A$ is a positive constant. In particular, $\GG^{2\pm}_>(\alpha)=o(\alpha^{-2})$.

To address $\GG^{2\pm}_<$~\eqref{eqn:G2split msmall}, we will make use of Stirling's approximation~\eqref{eqn:Stirling expand}; however, we are unable to assume $m$ is fixed and must include $m$ in the error bound,
\begin{equation}\label{eqn:Stirling careful}
  \frac{(2k+1+|m|)!}{(2k+1-|m|)!}~=~\lr{2k+\frac{3}{2}}^{2|m|}\lr{1-\frac{|m|(4|m|^2-1)}{12\lr{2k+\frac{3}{2}}^2}+o(k^{-2})}\,,  
\end{equation}as $k\to\infty$ with $m$ bounded relative to $k$ such that $m^3k^{-1}\to0$.

We use the compact support of $\sigma$ to extend the limits of summation over $k$ in~\eqref{eqn:G2split msmall} to $0\leq k<\infty$. Elementary estimates, using $0<q<\tfrac12$ to guarantee that $m^3 k^{-1}\to0$ as $k\to\infty$, asymptotic formulae~\eqref{eqn:Stirling careful} and~\eqref{eqn:P expand} and the Euler-Maclaurin formula~\eqref{eqn:euler-mac}, give
\begin{equation}\label{eqn:G2expansion}
    \GG^{2\pm}_<(\alpha)~=~\sum_{m\in\mc{C}^{2\pm}}\frac{1}{2}\int_0^\infty\dd\omega\,n(\beta\omega)J^2_{|m|}(\omega R)\sigma(\pm\Gamma(\omega-m\Omega))+O(\alpha^{-2})\,,
\end{equation}where we have extended the summation over $m$ to $m\in\mc{C}^{2\pm}$ by using that in each term the integrand is exponentially suppressed such that the total sum of the added terms is exponentially suppressed for $m>M$.

\subsection{Combining $\GG^{1\pm}$ and $\GG^{2\pm}$}

The full expression for the leading and next-to-leading terms for $\GG(\alpha)$ is now obtained by combining $\GG^{1\pm}$~\eqref{eqn:G1expansion} and $\GG^{2\pm}$\eqref{eqn:G2expansion}, with the result
\begin{multline}\label{eqn:Gfirst expansion}
    \GG(\alpha)~=~\sum_{m=-\infty}^{\infty}\frac{1}{2}\int_0^\infty\dd\omega\,n(\beta\omega)J^2_{|m|}(\omega R)\lrb{\sigma(\Gamma(\omega-m\Omega))+\sigma(-\Gamma(\omega-m\Omega))}-\frac{R^2}{16\beta\alpha^2}\ln\lr{\frac\alpha R}\sigma(\Omega\Gamma)+O(\alpha^{-2})\,.
\end{multline}
To extract the response function from~\eqref{eqn:Gfirst expansion}, we rewrite the limits of the integral and the summation, perform a change of variables, and take the subleading term under the integral. This gives
\begin{multline}\label{eqn:G assume}
    \GG(\alpha)~=~\int_\RR\dd E\, E^{-2}\sigma(E)\bigg(\frac{E^2}{2\Gamma}\sum_{m>E/(\Gamma\Omega)}n(\beta\varpi_+)J_{|m|}^2(\varpi_+R)+\frac{E^2}{2\Gamma}\sum_{m>-E/(\Gamma\Omega)}n(\beta\varpi_+)J_{|m|}^2(\varpi_+R)\\
    -\frac{R^2E^2}{16\beta\alpha^2}\ln\lr{\frac\alpha R}\delta(E-\Omega\Gamma)\bigg)\, +O(\alpha^{-2})\,,
\end{multline}where $\varpi_\pm=m\Omega\mp E/\Gamma$. 

Finally, we recall that the result~\eqref{eqn:G assume} was obtained under the assumption $\supp\{\sigma\}\subset\RR_{>0}$. When this assumption is relaxed to allow $\supp\{\sigma\}\subset\RR\setminus\{0\}$,~\eqref{eqn:G assume} is replaced by
\begin{multline}\label{eqn:G final assume relax}
    \GG(\alpha)~=~\int_\RR\dd E\, E^{-2}\sigma(E)\bigg(\frac{E^2}{2\Gamma}\sum_{m>|E|/(\Gamma\Omega)}n(\beta\omega_+)J_{|m|}^2(\omega_+R)+\frac{E^2}{2\Gamma}\sum_{m>-|E|/(\Gamma\Omega)}n(\beta\omega_+)J_{|m|}^2(\omega_+R)\\
    -\frac{R^2E^2}{16\beta\alpha^2}\ln\lr{\frac\alpha R}\delta(|E|-\Omega\Gamma)\bigg)\, +O(\alpha^{-2})\,,
\end{multline}where $\omega_\pm=m\Omega\mp|E|/\Gamma$.

\section{Euclidean vacuum response at small cosmological constant}\label{app:integral expand}

In this Appendix, we verify the small-cosmological constant asymptotic form~\eqref{eqn:dS vacuum expansion} of the detector response in the Euclidean vacuum in RdS.

Let $a\in\RR\setminus\{0\}$, $0<V<1$, and $\eta>0$. We define
\begin{equation}\label{eqn:I eta}
    I(\eta)~\coloneq~\int_0^\infty\dd z\frac{\sin (az)}{z}\frac{g(\eta z)}{\sqrt{1-w^2(z)g^2(\eta z)}}\,,
\end{equation}where
\begin{subequations}
\begin{align}
    g(z)&~\coloneq~\frac{z}{\sinh(z)}\,,\\w(z)&~\coloneq~V\sinc(z)\,,
\end{align}    
\end{subequations}and the notation suppresses the dependence of $I(\eta)$ on $a$ and $V$. $I(\eta)$ is the integral that appears in~\eqref{eqn:vacuum new params} with $a=2ER(\Gamma_VV)^{-1}$. We shall find first two terms in the asymptotic expansion of $I(\eta)$ as $\eta\to0$ with $a$ and $V$ fixed. 

For the rest of this Appendix, we assume $a>0$. The outcome~\eqref{eqn:final eta integral} holds also for $a<0$ by parity.

To find the leading term, we split the integral~\eqref{eqn:I eta} as
\begin{equation}\label{eqn: int split}
    I(\eta)~=~\int_0^\infty\dd z\,\frac{\sin(az)}{z}g(\eta z)+\int_0^\infty\dd z\,\frac{\sin(az)}{z}\lrb{\frac{1}{\sqrt{1-w^2(z)g^2(\eta z)}}-1}g(\eta z)\,.
\end{equation}Using~\cite[(3.981.1)]{G+R}, the first integral in~\eqref{eqn: int split} evaluates to $\tfrac{\pi}{2}\tanh(\tfrac{a\pi}{2\eta})$, which tends to $\pi/2$ with corrections of order $O(\ee^{-a\pi/\eta})$ as $\eta\to0$. In the second integral in~\eqref{eqn: int split}, a dominated convergence argument allows us to take the limit under the integral. Combining, we find
\begin{equation}
    \lim_{\eta\to0}I(\eta)~=~\int_0^\infty\dd z\,\frac{\sin(az)}{z}
\frac{1}{\sqrt{1-w^2(z)}}\,,
\end{equation}where we have used the integral $\int_0^\infty\dd z\sin(az)/z=\pi/2$.

To find the next-to-leading term, we split the integral~\eqref{eqn:I eta} as
\begin{subequations}\label{eqn: list of ints}
    \begin{align}
        I(\eta)&~=~I_0+I_1+I_2+I_3\,,\\
        I_0&~\coloneq~\int_0^\infty\dd z\,\frac{\sin(az)}{z}g(\eta z)\lrb{\frac{1}{\sqrt{1-w^2(z)}}-1-\frac{1}{2}w^2(z)}\,,\\
        I_1&~\coloneq~\int_0^\infty\dd z\,\frac{\sin(az)}{z}g(\eta z)\lrb{\frac{1}{\sqrt{1-w^2(z)g^2(\eta z)}}-\frac{1}{\sqrt{1-w^2(z)}}-\frac12w^2\lr{g^2(\eta z)-1}}\,,\\
        I_2&~\coloneq~\int_0^\infty\dd z\frac{\sin(az)}{z}g(\eta z)\,,\\
        I_3&~=~\frac12\int_0^\infty\dd z\,\frac{\sin(az)}{z}w^2(z)g^3(\eta z)\,.
    \end{align}
\end{subequations}

We consider first the integral $I_0$ in~\eqref{eqn: list of ints}. We introduce the function $h(x)=(g(x)-1)/x^2$, which is bounded and tends to $-\tfrac16$ as $x\to0$. Writing $g(\eta z)=1+\eta^2 z^2 h(\eta z)$ and splitting the integral gives
\begin{align}
    I_0~=~&\int_0^\infty\dd z\,\frac{\sin(az)}{z}\lrb{\frac{1}{\sqrt{1-w^2(z)}}-1-\frac{1}{2}w^2(z)}\nonumber\\
    &+\eta^2\int_0^\infty\dd z\,\sin(az)zh(\eta z)\lrb{\frac{1}{\sqrt{1-w^2(z)}}-1-\frac{1}{2}w^2(z)} \nonumber\\
    ~=~&\int_0^\infty\dd z\,\frac{\sin(az)}{z}\lrb{\frac{1}{\sqrt{1-w^2(z)}}-1-\frac{1}{2}w^2(z)}\nonumber\\
    &-\frac16\eta^2\int_0^\infty\dd z\,\sin(az)z\lrb{\frac{1}{\sqrt{1-w^2(z)}}-1-\frac{1}{2}w^2(z)}+o(\eta^2)\,,\label{eqn:I0 eval}
\end{align}where the second equality holds by a dominated convergence argument.

We consider next the integral $I_1$ in~\eqref{eqn: list of ints}. $I_1$ can be rearranged as
\begin{align}
    I_1~=~&\eta^2\int_0^\infty\dd z\,\sin(az)w^2(z)g(\eta z)\lr{g(\eta z)+1}h(\eta z)\nonumber\\&\times\lrb{\frac{1}{\sqrt{1-w^2(z)}\sqrt{1-w^2(z)g^2(\eta z)}\lr{\sqrt{1-w^2(z)}+\sqrt{1-w^2(z)g^2(\eta z)}}}-\frac{1}{2}}\nonumber \\~=~&-\frac16\eta^2\int_0^\infty\dd z\sin(az)zw^2(z)\lrb{\frac{1}{\lr{1-w^2(z)}^{3/2}}-1}+o(\eta^2)\,,\label{eqn:I1 eval}
\end{align}
where we again used $g(\eta z)=1+\eta^2 z^2h(\eta z)$, 
and in the second equality we have taken the limit under the integral by a dominated convergence argument.

We consider next the integral $I_2$ in~\eqref{eqn: list of ints}. Using~\cite[(3.981.1)]{G+R}, the integral $I_2$ evaluates as
\begin{align}
    I_2&~=~\frac{\pi}{2}\tanh\lr{\frac{a\pi}{2\eta}}\,\nonumber\\
    &~=~\frac\pi2+O(\ee^{-a\pi/\eta})\,.\label{eqn:I2 eval}
\end{align}

We consider next the integral $I_3$ in~\eqref{eqn: list of ints}. We extend the integral over the full real line by parity, perform a change of variables $z=u/\eta$, and deform the contour to $u=\ii\tfrac{\pi}{2}+r$ with $r\in\RR$, leading to
\begin{align}
    I_3~=~&\frac{1}{16}V^2\eta^2\int_{-\infty}^\infty
\dd r\frac{1}{\cosh^3 \! r}\bigg[\sinh\lr{\frac{\pi(a+2)}{2\eta}}\cos\lr{\frac{a+2}{\eta}r}
+\sinh\lr{\frac{\pi(a-2)}{2\eta}}\cos\lr{\frac{a-2}{\eta}r}\nonumber\\&\hphantom{abcdefghijklmnpoqwrsabcdef}-2\sinh\lr{\frac{a\pi}{2\eta}}\cos\lr{\frac{a}{\eta}r}\bigg] \nonumber\\
~=~&\frac{\pi}{32}V^2\lrb{\lr{(a+2)^2+\eta^2}\tanh\lr{\frac{\pi(a+2)}{2\eta}}+\lr{(a-2)^2+\eta^2}\tanh\lr{\frac{\pi(a-2)}{2\eta}}-2(a^2+\eta^2)\tanh\lr{\frac{a\pi}{2\eta}}} \nonumber\\
~=~&\frac{\pi}{16}V^2\lrb{4-\lr{(a-2)^2+\eta^2}\Theta(2-a)}+o(\eta^2)\,,\label{eqn:I3 eval}
\end{align}where $\Theta$ is the Heaviside theta function with the convention $\Theta(0)=1/2$. In the second equality, we have used~\cite[(3.985.1)]{G+R} and~\cite[(5.4.4)]{NIST} to evaluate the integrals. In the third equality, we have performed the elementary expansion as $\eta\to0$.

Combining~\eqref{eqn:I0 eval}-\eqref{eqn:I3 eval}, we have
\begin{align}
    I(\eta)~=~&\int_0^\infty\dd z\,\frac{\sin(az)}{z}
    \frac{1}{\sqrt{1-w^2(z)}}\nonumber\\
    &+\eta^2\lr{-\frac16\int_0^\infty\dd z \,z\sin(az)\lrb{\frac{1}{\lr{1-w^2(z)}^{3/2}}-1-\frac32w^2(z)}-\frac{\pi}{16}V^2\Theta(2-a)}+o(\eta^2)\,.
\end{align}
The integral $\tfrac14 \int_0^\infty\dd z\, z\sin(az)w^2(z)$ in the coefficient of $\eta^2$ may be evaluated by an elementary contour integral and is found to exactly cancel the term involving $\Theta(2-a)$. Therefore, we have
\begin{equation}\label{eqn:final eta integral}
    I(\eta)~=~\int_0^\infty \dd z\,\frac{\sin(a z)}{z}\frac{1}{\sqrt{1-w^2(z)}}-\frac16\eta^2\int_0^\infty\dd z\,z\sin(az)\lrb{\frac{1}{\lr{1-w^2(z)}^{3/2}}-1}+o(\eta^2)\,.
\end{equation}

\twocolumngrid
\bibliography{zz_bibliography}

\end{document}